\documentclass[useAMS,usenatbib]{mn2e}

\usepackage{graphicx}
\newcommand{\MS}{\ifmmode{\,}\else\thinspace\fi{\rm M}\ifmmode_{\odot}\else$_{\odot}$\fi}
\newcommand{\LS}{\ifmmode{\,}\else\thinspace\fi{\rm L}\ifmmode_{\odot}\else$_{\odot}$\fi}
\newcommand{\RS}{\ifmmode{\,}\else\thinspace\fi{\rm R}\ifmmode_{\odot}\else$_{\odot}$\fi}
\newcommand{\Ke}{\ifmmode{\,}\else\thinspace\fi{\rm K}}

\newcommand{\teff}{\ifmmode T_{\rm eff}\else$T_{\rm eff}$\fi}

\newcommand{\setF}{S1}%A
\newcommand{\setB}{S2}%B
\newcommand{\setA}{S3}%C
\newcommand{\setH}{S4}%D

\newcommand{\BEPRR}{\mbox{RRLYR-02792}}
\newcommand{\BEP}{\mbox{the BEP}}
\title[Pulsation models for binary evolution pulsators]{Pulsation models for the 0.26\MS\ star mimicking RR~Lyrae pulsator. Model survey for the new class of variable stars}
\author[Smolec et al.]
{R. Smolec$^{1}$\thanks{E-mail:
smolec@camk.edu.pl}, 
G. Pietrzy\'nski$^{2,3}$, 
D. Graczyk$^{2}$, 
B. Pilecki$^{2,3}$, 
W. Gieren$^{2}$, 
\and
I. Thompson$^{4}$, 
K. St\c{e}pie\'n$^{3}$,
P. Karczmarek$^{3}$, 
P. Konorski$^{3}$, 
M. G\'orski$^{3}$, 
\and
K. Suchomska$^{3}$,
G. Bono$^{5,6}$, 
P.G. Prada Moroni$^{7,8}$, 
N. Nardetto$^{9}$\\
$^{1}$Copernicus Astronomical Centre, Polish Academy of Sciences, Bartycka 18, 00-716 Warsaw, Poland\\
$^{2}$Departamento de Astronom\`ia, Universidad de Concepci\`on, Casilla 160-C, Concepci\`on, Chile\\
$^{3}$Warsaw University Observatory, Al. Ujazdowskie 4, 00-478 Warsaw, Poland\\
$^{4}$Carnegie Observatories, 813 Santa Barbara Street, Pasadena, CA 91101-1292, USA\\
$^{5}$Dipartimento di Fisica Universit\`a di Roma Tor Vergata, via della Ricerca Scientifica 1, 00133 Rome, Italy\\
$^{6}$INAF - Osservatorio Astronomico di Roma, Via Frascati 33, 00040 Monte Porzio Catone, Italy\\
$^{7}$Dipartimento di Fisica `E. Fermi', Universit\`a di Pisa, Largo B. Pontecorvo, 3, I-56127, Pisa, Italy\\
$^{8}$INFN, Sezione di Pisa, Largo B. Pontecorvo, 3, I-56127, Pisa, Italy\\
$^{9}$Laboratoire Lagrange, UMR7293, UNS/CNRS/OCA, 06300 Nice, France
}

\begin{document}

\date{Accepted . Received ; in original form }

\pagerange{\pageref{firstpage}--\pageref{lastpage}} \pubyear{2011}

\maketitle

\label{firstpage}

\begin{abstract}
We present non-linear hydrodynamic pulsation models for OGLE-BLG-RRLYR-02792 -- a $0.26\MS$ pulsator, component of the eclipsing binary system, analysed recently by Pietrzy\'nski et al. The star's light and radial velocity curves mimic that of classical RR~Lyrae stars, except for the bump in the middle of the ascending branch of the radial velocity curve. We show that the bump is caused by the 2:1 resonance between the fundamental mode and the second overtone -- the same mechanism that causes the Hertzsprung bump progression in classical Cepheids. The models allow to constrain the parameters of the star, in particular to estimate its absolute luminosity ($\approx 33\LS$) and effective temperature ($\approx 6970$\thinspace K, close to the blue edge of the instability strip). 

We conduct a model survey for the new class of low mass pulsators similar to OGLE-BLG-RRLYR-02792 -- products of evolution in the binary systems. We compute a grid of models with masses corresponding to half (and less) of the typical mass of RR~Lyrae variable, $0.20\MS\le M\le 0.30\MS$, and discuss the properties of the resulting light and radial velocity curves. Resonant bump progression is clear and may be used to distinguish such stars from classical RR~Lyrae stars. We present the Fourier decomposition parameters for the modelled light and radial velocity curves. The expected values of the $\varphi_{31}$ Fourier phase for the light curves differ significantly from that observed in RR~Lyrae stars, which is another discriminant of the new class.
\end{abstract}


%%%%%%%%%%%%%%%%%%%%%%%%%%%%%%%%%%%%%%%
\section{Introduction}\label{sec.intro}
%%%%%%%%%%%%%%%%%%%%%%%%%%%%%%%%%%%%%%%

\begin{figure}
\centering
\resizebox{\hsize}{!}{\includegraphics{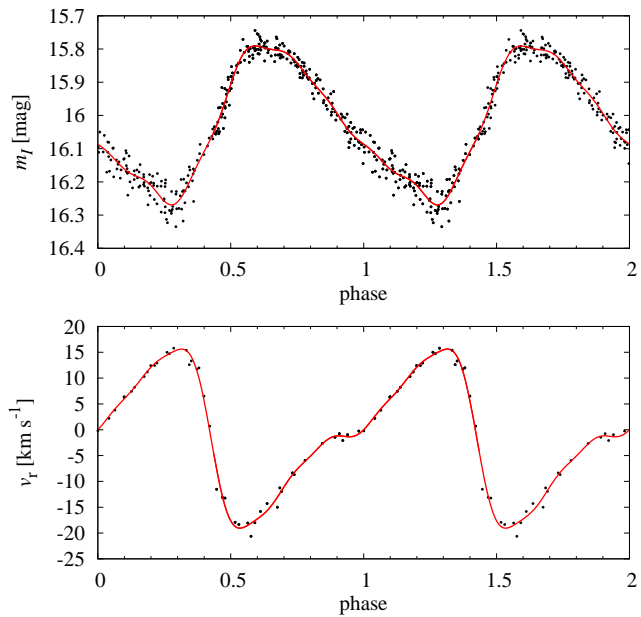}}
\caption{Disentangled $I$-band light curve (top) and radial velocity curve (bottom) for \BEPRR\ (the BEP). Solid lines represent the 10th (light curve) and 6th (radial velocity curve) order Fourier fits to the data. Fourier decomposition parameters are given in Tab.~\ref{tab.fourier}. Note that we use an observers' convention in the plot of radial velocity curve, i.e. negative values correspond to expansion.}
\label{fig.lrvc}
\end{figure}

\begin{figure}
\centering
\resizebox{\hsize}{!}{\includegraphics{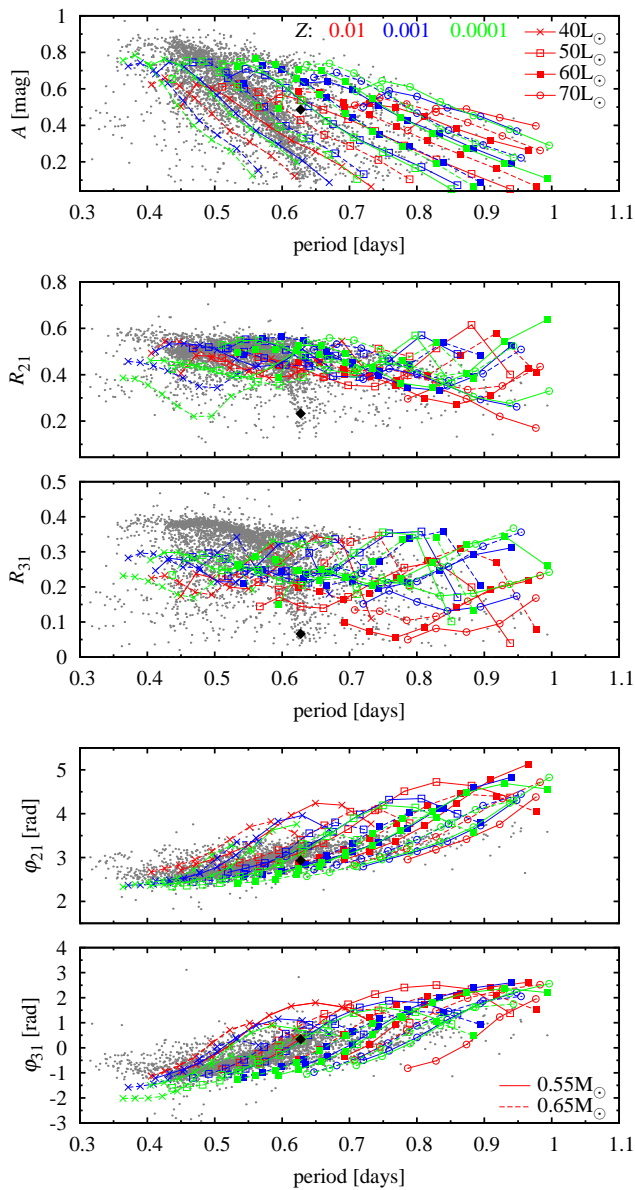}}
\caption{Peak-to-peak amplitude, $A$, and Fourier decomposition parameters for the $I$-band light curves of the Galactic bulge RR~Lyrae stars of the OGLE survey \citep[grey dots, each third plotted;][]{sosz_rrl_blg}, \BEPRR\ (the BEP, black diamond) and hydrodynamic models adopting convective parameters of set \setF\ (lines, see the keys in the top and bottom panels).}
\label{fig.fou_normF}
\end{figure}

OGLE-BLG-RRLYR-02792 (below shortened to RR) is a large amplitude, single-periodic ($P_0\approx 0.6275$\thinspace d) pulsator, component of a physical, well detached, double-lined eclipsing binary system discovered in the Galactic bulge by the Optical Gravitational Lensing Experiment (OGLE) survey \citep{sosz_rrl_blg} and analysed in detail in the course of the Araucaria project by \cite{gp12}. Disentangled  $I$-band light curve and radial velocity curve are presented in Fig.~\ref{fig.lrvc}. The pulsation period and shape of the light curve places the star among classical RR~Lyrae stars pulsating in the fundamental mode (RRab stars), which is clear from the analysis of the Fourier decomposition parameters, amplitude ratios, $R_{k1}$, and Fourier phases, $\varphi_{k1}$. They are constructed through the fit of the Fourier series to the data, i.e.,
\begin{equation}
m_I=A_0+\sum_{k}A_k\sin(k\omega t+\varphi_k),
\end{equation}
and then,
\begin{equation}
R_{k1}=A_k/A_1,\ \ \varphi_{k1}=\varphi_k-k\varphi_1,
\end{equation}
where $\omega=2\pi/P$ is pulsation frequency. As computed by \cite{gp12} the pulsation period decreases with the rate of $(8.4\pm 2.6)\times 10^{-6}$\thinspace d\thinspace yr$^{-1}$. Here we adopt the mean pulsation period derived from the $I$-band photometry, which spans nearly 10\thinspace years, $P_0=0.627527$\thinspace d. Fourier decomposition parameters of \BEPRR\ are given in Tab.~\ref{tab.fourier} and plotted in Fig.~\ref{fig.fou_normF} along with the OGLE data for thousands of RRab stars from the Galactic bulge \citep{sosz_rrl_blg}. Note that the Fourier phases from the OGLE catalogue, given in the cosine convention, were shifted to match the sine convention (eq. 1) that we consistently use in this paper (for both the light and radial velocity curves)\footnote{$\varphi_{21}^{\rm (sine)}=\varphi_{21}^{\rm (cosine)}-\pi/2$,\\ $\varphi_{31}^{\rm (sine)}=\varphi_{31}^{\rm (cosine)}-\pi$.}. Fourier phases of \BEPRR\ fit the typical values of RR~Lyrae stars very well. Amplitude ratios are low, but fit the tail well visible in the progression for RRab stars between $P_0=0.6$ and $0.65$\thinspace days. 

Surprisingly, the dynamical mass of \BEPRR\ derived by \cite{gp12} is very low, $M=0.261\pm0.015$\thinspace\MS, which is roughly two times smaller than the typical mass of the RR~Lyrae star. It is too low to initiate the helium burning. As discussed by \cite{gp12} the star is a result of mass transfer in the binary system. Other parameters of \BEPRR\ relevant for this study and derived by \cite{gp12} are radius, $R=4.24\pm 0.24\RS$ and estimate of the effective temperature, $\teff=7320\pm 160$\thinspace K (rough estimate based on the assumed $5000$\thinspace K temperature of the secondary $1.67\MS$ giant component and temperature ratio derived from the analysis of the light curve). Corresponding absolute luminosity may be estimated to be $L=46.2\pm 9.3$\LS. No metallicity determination is available for \BEPRR. 

\BEPRR\ is a new member of class of pulsators, for which mass transfer associated with binary evolution has significant effect on their pulsation properties. We will call such stars binary evolution pulsators (BEPs in the following) and we will call the \BEPRR\ the BEP. Other well known examples of the class are hot subdwarf pulsators, members of eclipsing binary systems \citep[see][]{for}. We also note that BEP may be a single object that went through binary evolution in the past, and was ejected from the binary because of e.g. tidal interactions in the globular cluster.

\begin{table}
\caption{Fourier decomposition parameters for the $I$-band light curve and radial velocity (RV) curve of \BEPRR\ (the BEP, sine decomposition). Standard errors are given in the parentheses. In the last line the peak-to-peak amplitude, $A$, is given.}
\label{tab.fourier}
\centering
\begin{tabular}{lll}
\hline
              & $I$-band & RV\\
\hline
$A_1$ [mag$|$km\thinspace s$^{-1}$]& 0.218(2)  & 13.46(27)\\
$R_{21}$                 & 0.232(12) & 0.526(27)\\
$R_{31}$                 & 0.066(11) & 0.160(21)\\
$\varphi_{21}$ [rad]     & 2.93(5)   & 3.11(6)\\
$\varphi_{31}$ [rad]     & 0.34(17)  & -0.17(15)\\
$A$  [mag$|$km\thinspace s$^{-1}$] & 0.486     & 34.67\\
\hline
\end{tabular}
\end{table}

In this paper we present the first pulsation calculations for \BEP, both linear and non-linear, full-amplitude models. Nearly exact fit of the light and radial velocity curves is possible, which allows to constrain the stellar parameters and provides an insight into the pulsation dynamics of the star (Section~\ref{sec.resultsBEP}). We also present a non-linear model survey with parameters of the models corresponding to the expected parameters of the members of the new class, i.e. with masses $2-3$ times smaller than typically assumed for RR~Lyrae stars (Section~\ref{sec.survey}). The main goal is to provide the criteria to distinguish the BEPs from classical RR~Lyrae stars, for which binary nature cannot be inferred directly because of too low inclination angle. In this case, we only see the light/radial velocity curve of a pulsating component, which, as in the case of \BEP, may strongly mimic the typical curve of RR~Lyrae pulsator.

%%%%%%%%%%%%%%%%%%%%%%%%%%%%%%%%%%%%%%%%%%%%%%
\section{Numerical methods}\label{sec.hydro}
%%%%%%%%%%%%%%%%%%%%%%%%%%%%%%%%%%%%%%%%%%%%%%

All our models are computed with the convective pulsation codes described in detail by \cite{sm08a}. These are static model builder, linear non-adiabatic code and non-linear direct time-integration hydrodynamic code. The codes are Lagrangian. Radiative transfer is treated with a simple diffusion approximation. For the convective energy transport we use the time-dependent, turbulent convection model of \cite{ku86}. It is a simple one-dimensional, one-equation formulation that contains several dimensionless scaling parameters. These are the mixing-length parameter, $\alpha$, and parameters scaling the turbulent fluxes and the terms that drive/damp the turbulent energy: $\alpha_{\rm p}$ (turbulent pressure), $\alpha_{\rm m}$ (eddy-viscous dissipation), $\alpha_{\rm c}$ (convective heat flux), $\alpha_{\rm t}$ (kinetic turbulent energy flux), $\alpha_{\rm s}$ (buoyant driving), $\alpha_{\rm d}$ (turbulent dissipation) and $\gamma_{\rm r}$ (radiative cooling). Theory provides no guidance for their values. For $\alpha_{\rm s}$, $\alpha_{\rm c}$, $\alpha_{\rm d}$ and $\gamma_{\rm r}$ a values which result from comparison of the static time-independent version of the model with the Mixing Length Theory (MLT) are in use (see caption of Tab.~\ref{tab.convpar}). In practice, the values of convective parameters are calibrated to match as many observational constraints as possible. Slightly different values are necessary for modelling different groups of stars, e.g. Cepheids and RR~Lyrae stars of different stellar systems. It makes the modelling of BEPs particularly difficult. The only known member of the class, \BEP, provides very little constraints on the models. Consequently, several sets of values of convective parameters need to be considered -- Tab.~\ref{tab.convpar}. As a first guess we adopt the parameters that well reproduce the observed properties of the Galactic bulge RR~Lyrae stars, in particular the Fourier parameters of the $I$-band light curves. In Fig.~\ref{fig.fou_normF} we compare the computed Fourier parameters for models adopting convective parameters of set \setF, and assuming typical parameters of RR~Lyrae stars (masses $0.55\MS$ and $0.65\MS$, luminosities: $40-70\LS$, metallicities, $Z$: $0.01$, $0.001$, and $0.0001$) with OGLE observations. We get a good agreement for amplitudes, Fourier phases and $R_{21}$ ratio and somewhat too low model values for $R_{31}$. Similar agreement (with much better match of $R_{31}$) is obtained for models of set \setB, which include the effects of radiative cooling. In set \setA\ the convective parameters are the same as in set \setF, except eddy viscosity which is lower. Consequently, we expect higher pulsation amplitudes in models adopting parameters of set \setA. In linear model survey we also consider set \setH, in which effects of turbulent pressure are turned on. Inclusion of turbulent pressure in non-linear models leads to long-lasting integration and convergence difficulties for some models. Consequently, in this first non-linear model survey for BEPs, we neglect the effects of turbulent pressure and consider models of sets \setF, \setB\ and \setA, only. 

Our models are chemically homogeneous, non-rotating, envelope models. They consist of $150$ mass shells extending down to $2\!\times\! 10^6$\thinspace K. It is a standard depth in modelling the Cepheid and RR~Lyrae pulsation. It is deep enough to cover the region relevant for $\kappa$ mechanism driven pulsation but too shallow to enter the inhomogeneous hydrogen burning shell region. In order to capture and resolve the thin hydrogen ionisation region the anchor zone of temperature equal to $11\,000$\thinspace K is located $40$ zones below the surface. Non-linear model integration is conducted till the limit cycle pulsation (single periodic, full-amplitude pulsation) is reached, which requires computation of $1\,000$ up to $5\,000$ pulsation cycles. Each pulsation cycle is covered with $600$ time-steps. The bolometric light curves are transformed to the $I$-band through the static \cite{kurucz} atmosphere models\footnote{http://kurucz.harvard.edu/}. Projection factor for \BEP\ is unknown. We scaled the model velocity curves by projection factor used for classical RR~Lyrae pulsators, i.e. $p=1.38$ \citep{fernley}.

The chemical structure of the star is subject of uncertainty. No metallicity determination is available. Also, no realistic evolutionary computations exist for this type of binary evolution. Significant fraction of the envelope was removed during the evolution. Pulsation is driven in the outermost layers which, although very extended, contain only a fraction of percent of the total stellar mass (typically below $1$\thinspace per cent in our models). In this study we assume that the outer shell has preserved the standard chemical composition typical for Population II stars, i.e., $X=0.76$, $Z=0.01-0.0001$. We also compute model sequences with $X=0.70$ which corresponds to higher helium content as in Population I stars. In all our models we adopt the OP opacities \citep{sea05} and solar mixture of \cite{a04}.   

\begin{table}
\caption{Convective parameters adopted in the models. Parameters $\alpha_{\rm s}$, $\alpha_{\rm c}$, $\alpha_{\rm d}$, $\alpha_{\rm p}$ and $\gamma_{\rm r}$ are given in the units of standard values \citep[$\alpha_{\rm s}=\alpha_{\rm c}=1/2\sqrt{2/3}$, $\alpha_{\rm d}=8/3\sqrt{2/3}$, $\alpha_{\rm p}=2/3$ and $\gamma_{\rm r}=2\sqrt{3}$; see][for details]{sm08a}. Only first three sets of parameters were used in non-linear computations.}
\label{tab.convpar}
\centering
\begin{tabular}{lcccccccc}
\hline
set & $\alpha$ & $\alpha_{\rm m}$ & $\alpha_{\rm s}$ & $\alpha_{\rm c}$ & $\alpha_{\rm d}$ & $\alpha_{\rm p}$ & $\alpha_{\rm t}$ & $\gamma_{\rm r}$ \\
\hline
\setF & 1.5 & 0.4 & 1.0 & 1.0 & 1.0 & 0.0 & 0.0  & 0.0\\
\setB & 1.5 & 0.6 & 1.0 & 1.0 & 1.0 & 0.0 & 0.0  & 1.0\\
\setA & 1.5 & 0.2 & 1.0 & 1.0 & 1.0 & 0.0 & 0.0  & 0.0\\
\hline
\setH & 1.5 & 0.6 & 1.0 & 1.0 & 1.0 & 1.0 & 0.01 & 0.0\\
\hline
\end{tabular}
\end{table}

%%%%%%%%%%%%%%%%%%%%%%%%%%%%%%%%%%%%%%%%%%%%%%%%%%%%%%%%%%
\section{Pulsation models for the BEP}\label{sec.resultsBEP}
%%%%%%%%%%%%%%%%%%%%%%%%%%%%%%%%%%%%%%%%%%%%%%%%%%%%%%%%%%

\begin{figure*}
\centering
\resizebox{\hsize}{!}{\includegraphics{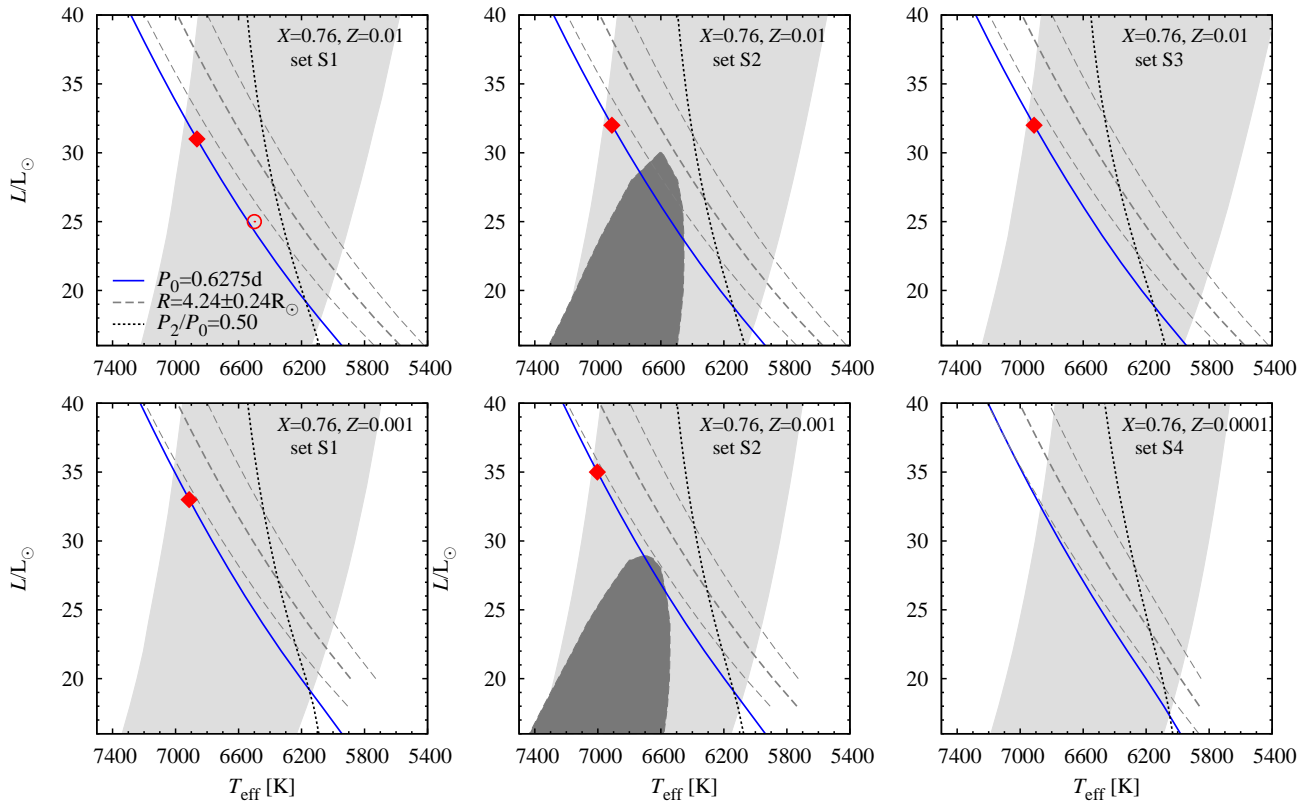}}
\caption{The HR diagrams for $0.26\MS$ models adopting different values of convective parameters and different chemical compositions as indicated in the top right part of each panel. Instability strips are indicated with gray-shaded areas (light-gray for the fundamental mode, dark-gray for the first overtone). Solid line is the line of constant fundamental mode period, $P_0=0.6275$\thinspace d. Long-dashed lines indicate the radius determination for \BEP\ (with 1$\sigma$ errors) and dotted line shows the loci of the 2:1 resonance with the second overtone, $P_2/P_0=0.5$. Circle marks location of the static model from Fig.~\ref{fig.struct}, and diamonds mark the location of the best matching models for \BEP\ (Section~\ref{ssec.nonlinear}, Tab.~\ref{tab.bestmodels}).}
\label{fig.hr}
\end{figure*}

In this Section we focus on modelling the observed properties of \BEP. We first conduct a linear model survey and discuss properties of the static models (Section~\ref{ssec.linear}). Next we focus on non-linear modelling of the $I$-band light curve and radial velocity curve of \BEP\ (Section~\ref{ssec.nonlinear}).

\subsection{Linear model survey}\label{ssec.linear}
%%%%%%%%%%%%%%%%%%%%%%%%%%%%%%%%%%%%%%%%%%%%%%%%%%%%%%

The models are computed along horizontal stripes of constant luminosity ($L=16-40\LS$, step $2\LS$) and with $100$\thinspace K step in effective temperature. We adopt $M=0.26\MS$ for all the models. Computed Hertzsprung-Russel (HR) diagrams for different sets of convective parameters and assuming different chemical compositions are plotted in Fig.~\ref{fig.hr}. Except for model computations adopting convective parameters of set \setB, first overtone is linearly damped in our models. The typical width of the fundamental mode instability strip is $1\,000$\thinspace K and increases towards the higher luminosities.

\begin{figure}
\centering
\resizebox{\hsize}{!}{\includegraphics{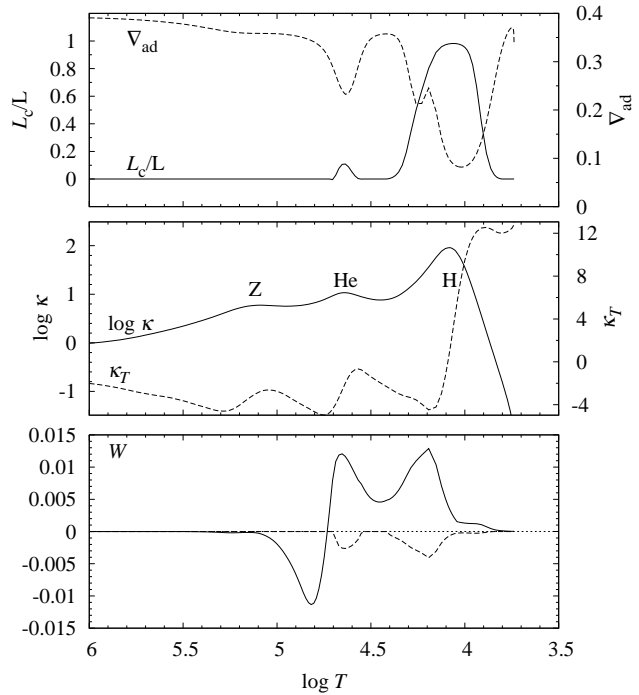}}
\caption{Properties of the exemplary static model with $M=0.26\MS$, $L=25\LS$ and $\teff=6500$\thinspace K (marked with circle in the top left panel of Fig.~\ref{fig.hr}). {\it Top panel:} convective heat flux (solid line) and adiabatic temperature gradient (dashed line); {\it middle panel:} logarithm of opacity (solid line) and its logarithmic temperature derivative (dashed line); {\it bottom panel:} work integrands, total (solid line) and eddy-viscous contribution (dashed line).}
\label{fig.struct}
\end{figure}

Before we discuss the models appropriate for \BEP, we first discuss general properties of typical static model, as pulsation properties of BEPs were not studied so far. In Fig.~\ref{fig.struct} we plot results for a model located in the middle of the instability strip ($\teff=6500$\thinspace K, $L=25\LS$, marked with open circle in the top left panel of Fig.~\ref{fig.hr}). Its period is only slightly longer than the period of \BEP. The top panel shows the convective heat flux (solid line) and adiabatic temperature gradient (dashed line). In the middle panel of Fig.~\ref{fig.struct} we plot the logarithm of Rosseland opacity (solid line) and its logarithmic temperature derivative (dashed line) -- crucial quantities for studying the pulsation driving. The metal, helium and hydrogen opacity bumps are marked with labels. In the bottom panel the work integrands are plotted -- total with the solid line and contribution of eddy viscous work with the dashed line. Pulsation driving occurs in regions with positive work integrand. Eddy viscosity always damps the pulsation and its contribution is negative. Model properties are very similar to the properties of classical RR~Lyrae models \citep[e.g.][]{stel82}. It is clear that pulsation is driven through the $\kappa$ mechanism working in the hydrogen-helium ionisation zone. We note a significant contribution from the hydrogen ionisation region. The main envelope convection zone is associated with the hydrogen-helium (HeI-H) ionisation region. The second, less effective convective region develops deeper at the HeII ionisation zone. All ionisation regions are properly resolved in our models as run of $\nabla_{\rm ad}$ indicates. 

In each panel of Fig.~\ref{fig.hr} we plot a line of constant fundamental mode period, $P_0=0.627527$\thinspace d. The dashed lines in Fig.~\ref{fig.hr} are lines of constant photospheric radius $R=4.24\RS$, with 1$\sigma$ errors ($\pm0.24\RS$), which correspond to the values measured for \BEP. For models with mass corresponding to the measured mass of \BEP, these two lines (of constant $P_0=0.6275$\thinspace d and of constant $R=4.24\RS$) should roughly overlap (period-mean density relation), which is not the case for any of the considered model parameters. The lines are nearly parallel but shifted with respect to each other. The best agreement is obtained for models of the lowest metal abundance and for the models including the effects of turbulent pressure, which expands the model envelope (middle bottom panel in Fig.~\ref{fig.hr}). In this case the difference is 1$\sigma$ (at high $L$). For other models the difference is larger, however, always smaller than $2\sigma$. Hence, we treat this discrepancy as a warning sign and note that multi-band photometric observations are necessary to derive the radius of \BEP\ more accurately (see also discussion in Section~\ref{sec.concl}).

In the HR diagrams in Fig.~\ref{fig.hr} we also plot the loci of the 2:1 resonance between the fundamental mode and the second overtone, $P_2/P_0=0.5$ (dotted lines). This resonance causes the famous Hertzsprung bump progression in classical Cepheids pulsating in the fundamental mode \citep[e.g.][]{ss76,kovb89,bmk90}. In the discussed models it lays well within the instability strip, independent of the model parameters (see also Fig.~\ref{fig.hr2} for models with different mass). Our non-linear computations presented in the next sections show that this resonance strongly influences the pulsation of BEPs.

\subsection{Non-linear model survey}\label{ssec.nonlinear}
%%%%%%%%%%%%%%%%%%%%%%%%%%%%%%%%%%%%%%%%%%%%%%%%%%%%%%%%%%%%

\begin{figure*}
\centering
\resizebox{.49\hsize}{!}{\includegraphics{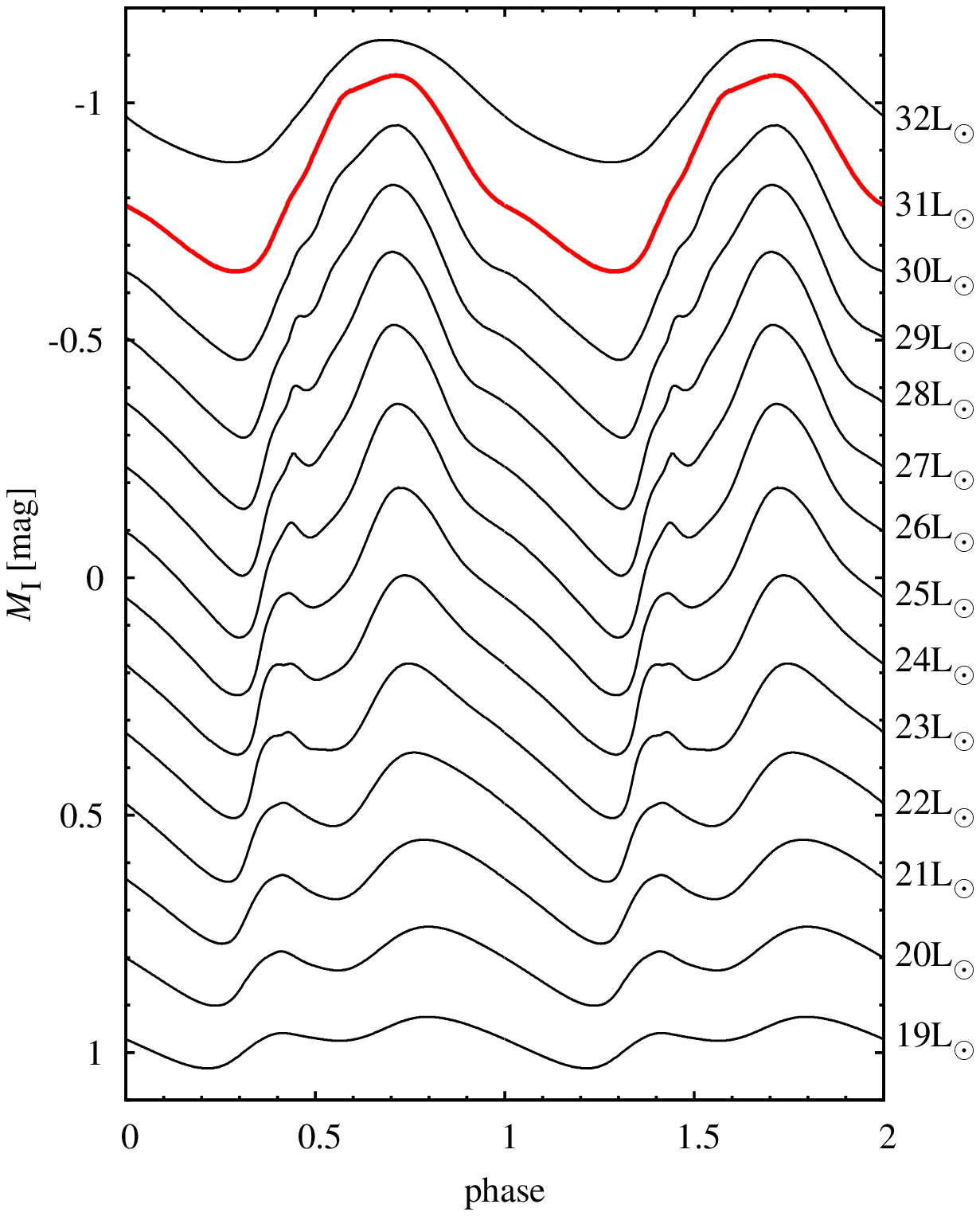}}
\resizebox{.49\hsize}{!}{\includegraphics{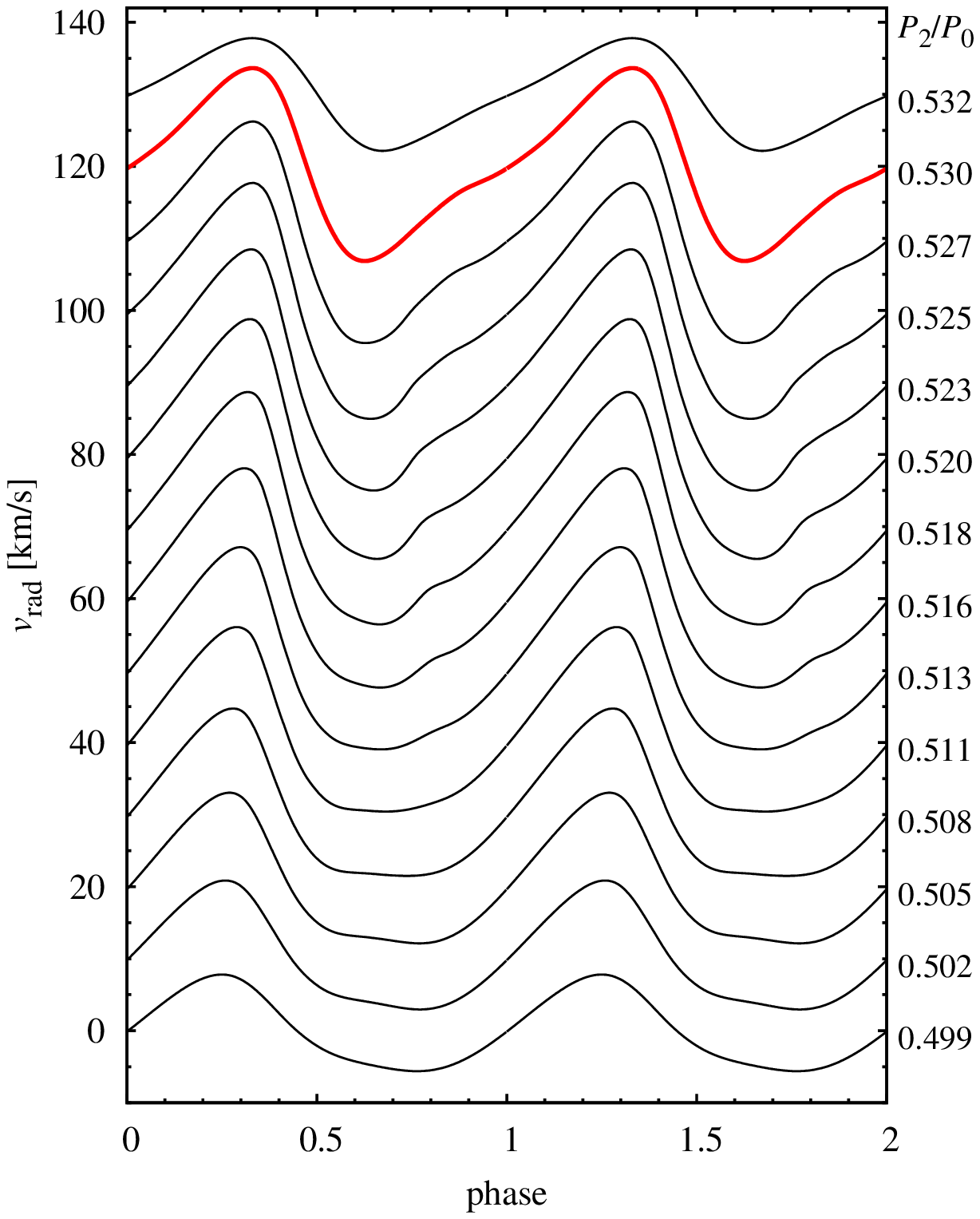}}
\caption{$I$-band light curves (left panel) and radial velocity curves (right panel, observers convention, i.e. negative values correspond to envelope expansion) for hydrodynamic models of $0.26\MS$ computed along a line of constant fundamental mode period corresponding to the pulsation period of \BEP. Models adopt convective parameters of set \setF. For clarity, the consecutive light curves were shifted by $0.12$\thinspace mag and radial velocity curves by $10$\thinspace km\thinspace s$^{-1}$. Velocity curves were scaled by constant projection factor, $p=1.38$. Thick red lines indicate the best matching model for \BEP.}
\label{fig.PER.curves}
\end{figure*}

The non-linear models for \BEP\ are computed along lines of constant fundamental mode period, $P_0=0.6275$\thinspace d. Models are labelled by their absolute luminosity, which increases in a step of $1\LS$. All models have $M=0.26\MS$. The least luminous models are located at the red edge of the instability domain and, as luminosity increases, shift towards the blue (see Fig.~\ref{fig.hr}). We computed six model sequences: with convective parameters of sets \setF, \setB\ and \setA\ and with metallicities $Z=0.01$ and $Z=0.001$ ($X=0.76$) in each case. For models with convective parameters of set \setA\ we also computed two model sequences with $X=0.70$ ($Z=0.01$ and $Z=0.001$). In Fig.~\ref{fig.PER.curves} we show a full set of the computed $I$-band light curves and radial velocity curves for models with convective parameters of set \setF\ and with $Z=0.01$. At phase $0$ the radius reaches its maximum value (velocity is equal to 0). The radial velocity curves have a usual triangular shape. One may note a systematic progression of the curve's shape as luminosity changes. For the lowest luminosity models the velocity curves are more or less symmetrical, the descending branch is even slightly longer than the ascending branch for the  model  with $L=19\LS$. The velocity varies slowly at the phase of maximum expansion velocities. For models with $L=22-23\LS$ the velocity curve is nearly flat at this phase. As luminosity increases a bump appears on the ascending branch (visible already for $L=25\LS$ at phase around $0.8$), which moves upwards towards the phase of maximum radius ($L=31\LS$) and then, vanish ($L=32\LS$).

Considering the $I$-band light curves, at the lowest luminosities they differ significantly from the typical RR~Lyrae curves. A pronounced bump is present on the ascending branch -- two brightness maxima are observed during one pulsation cycle. As luminosity increases the bump becomes less pronounced, turns into narrow standstill ($28,\ 29 \LS$) and vanishes. At highest luminosities, the light curves resemble those of typical RR~Lyrae stars, with the except for less sharp light variation in the brightness maximum.

Described variation of the shape of radial velocity and $I$-band light curves suggests, that the 2:1 resonance with the second overtone is influential here, and shapes the bump progression. It is not very pronounced because we consider a specific model sequences of constant period. It is clear form Fig.~\ref{fig.hr} that most of our models lay to the blue of the $P_2/P_0=0.5$ line and thus, majority of models have $P_2/P_0>0.5$ -- see exact values on the right side of Fig.~\ref{fig.PER.curves}. In the case of bump progression observed e.g. in classical Cepheids, which is caused by the same resonance, the shape of the light/radial velocity curves is affected by the resonance in a wide range of period ratios, $0.48<P_2/P_0<0.54$. The position of the bump depends critically on the value of $P_2/P_0$, as model computations indicate \citep{kovb89,bmk90}. For $P_2/P_0>0.5$ the bump is located on the ascending branch of the velocity curve, which is also the case in our models. The full bump progression in our models will be demonstrated in Section~\ref{sec.survey}, where we consider models covering the full instability strip and thus probing the $P_2/P_0$ ratio in a wide range (Fig.~\ref{fig.HP_curves}).

\begin{figure}
\centering
\resizebox{\hsize}{!}{\includegraphics{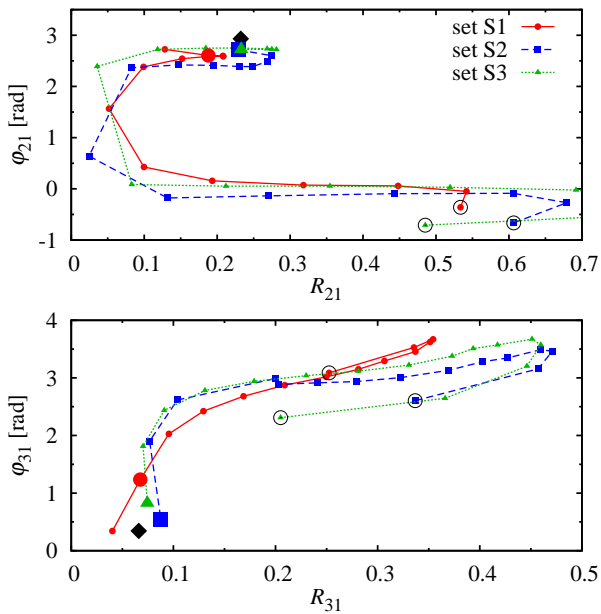}}
\caption{Plots of the Fourier parameters, $\varphi_{21}$ vs. $R_{21}$ (top) and $\varphi_{31}$ vs. $R_{31}$ (bottom) for the computed $I$-band light curves for three sequences of models of constant period (all models with $X=0.76$ and $Z=0.01$, but with different convective parameters, as indicated in the key). Open circle marks the first, lowest luminosity model in each sequence. Consecutive models differ by $1\LS$. Larger symbols show the best matching models for \BEP\ (marked with diamond).}
\label{fig.best_lc}
\end{figure}

\begin{figure}
\centering
\resizebox{\hsize}{!}{\includegraphics{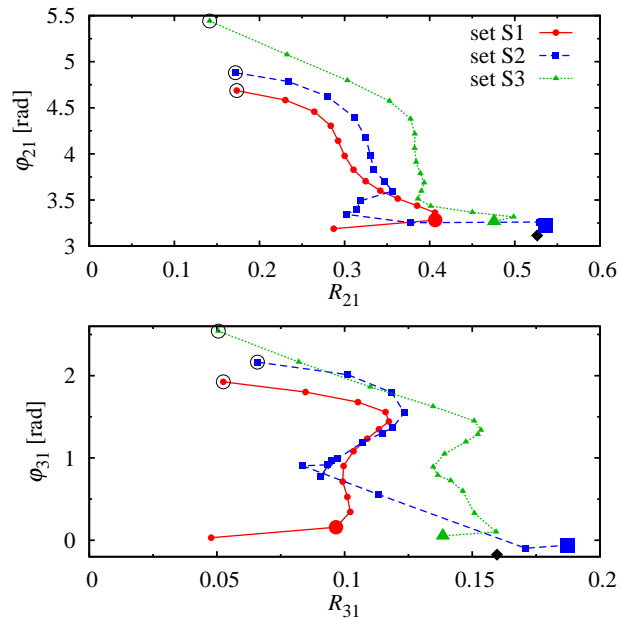}}
\caption{The same as Fig.~\ref{fig.best_lc}, but for the radial velocity curves.}
\label{fig.best_rv}
\end{figure}

It is clear that among the models displayed in Fig.~\ref{fig.PER.curves}, these are the highest luminosity ones that reproduce the observations of \BEP\ best (cf. Fig.~\ref{fig.lrvc}). In particular, the model for $L=31\LS$ (marked with thick red lines in Fig.~\ref{fig.PER.curves}) seems to be the best one. A more quantitative selection of the best models for \BEP\ is based on comparison of the observed and computed Fourier decomposition parameters of the lowest order, the amplitude $A_1$, amplitude ratios, $R_{21}$ and $R_{31}$, and Fourier phases, $\varphi_{21}$ and $\varphi_{31}$, for both $I$-band and radial velocity curves. In Figs.~\ref{fig.best_lc} and \ref{fig.best_rv} we show such comparisons for specific Fourier parameter planes, i.e., $\varphi_{21}-R_{21}$ and  $\varphi_{31}-R_{31}$ and three sequences of models. We stress that periods of all the models match the period of \BEP. Models differ in luminosity (and effective temperature) -- the least luminous model (and the coolest one) is marked with open circle. Luminosity of the consecutive models in a sequence differs by $1\LS$. Again it is clear that these are the highest luminosity models that match the observations of \BEP\ best. 

Yet another observable that can be compared with the models is the phase lag between the radial velocity curve and the light curve, defined as $\Delta\varphi_1=\varphi_1^{\rm vel}-\varphi_1^{\rm mag}$  \citep{simon84}. For \BEP\ we get $\Delta\varphi_1=0.055\pm 0.027$\footnote{ Because of the pulsation period change, $\Delta \varphi_1$ was computed using the last 500\thinspace days of the OGLE-III photometry only. At the end of this span the radial velocities were measured. The adopted pulsation period ($P=0.627510(5)$\thinspace d) was derived from the photometry alone, but well agrees with the (less accurate) period derived from radial velocity measurements.}. Unfortunately there are no phase lag determinations for classical RR~Lyrae stars using the $I$-band light curves. With the $V$-band light curves, the mean phase lag for a sample of RR Lyrae stars is $\langle\Delta\varphi_1\rangle=-0.19$ \citep{omk00}. It is negative for all the stars used in the work of \cite{omk00} (Moskalik private comm.). Our models for classical RRab stars (displayed in Fig.~\ref{fig.fou_normF}) point however, that phase lags computed using the $I$-band light curves are systematically larger than phase lags computed using the $V$-band light curves and may have both positive and negative values.

We now compute the offsets between the model and the observations in the Fourier parameter space, defined as
\begin{equation}
d=\sqrt{\sum_i(p_{i, \rm{mod}}-p_{i,\rm{obs}})^2/p_{i,\rm{obs}}^2},
\end{equation}
where $p_i$ denotes one of the low order Fourier parameters and amplitude, $p_i\in\{A, R_{21}, R_{31}, \varphi_{21}, \varphi_{31}\}$, for our models ($p_{i, \rm{mod}}$) and for the observed curves ($p_{i,\rm{obs}}$, Tab.~\ref{tab.fourier}). Depending on the Fourier parameters used, several offsets may be constructed. At first we consider three definitions of $d$. The first offset, $d_2$, is computed in the 2D-plane of the lowest order Fourier coefficients, $\varphi_{21}-R_{21}$ (top panels of Figs.~\ref{fig.best_lc} and \ref{fig.best_rv}). The second offset, $d_4$, is constructed of all four low order amplitude ratios and Fourier phases, and the third offset, $d_5$, includes also the amplitude. 

We note that for a given model sequence and for both the $I$-band and radial velocity curves the three offsets are minimised either for the same model or by models that differ by $1\LS$. For example, for models adopting convective parameters of set \setB, and $Z=0.01$ all offsets for the $I$-band light curves and radial velocity curves are minimised in the model with $L=32\LS$, except $d_4$ for the radial velocity curves which is minimised in the model with $L=31\LS$. In the more complex case of models with convective parameters of set \setF\ (and $Z=0.01$) $d_4$ and $d_5$ for the $I$-band light curves and $d_2$ for the radial velocity curves are minimised in the model with $L=31\LS$, while $d_2$ for the $I$-band light curves and $d_4$ and $d_5$ for the radial velocity curves are minimised in the model with $L=30\LS$.

\begin{figure*}
\centering
\resizebox{.49\hsize}{!}{\includegraphics{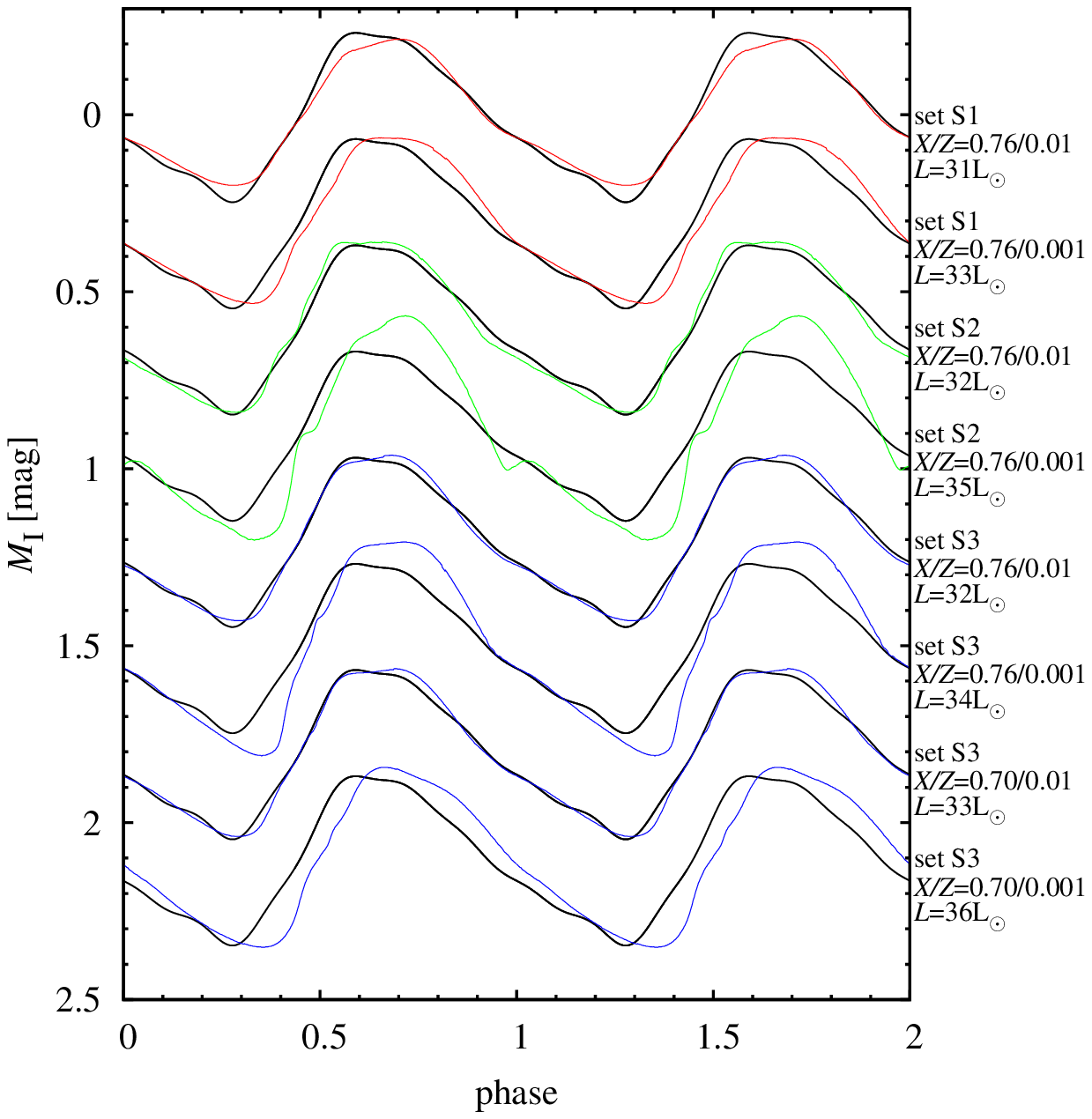}}
\resizebox{.49\hsize}{!}{\includegraphics{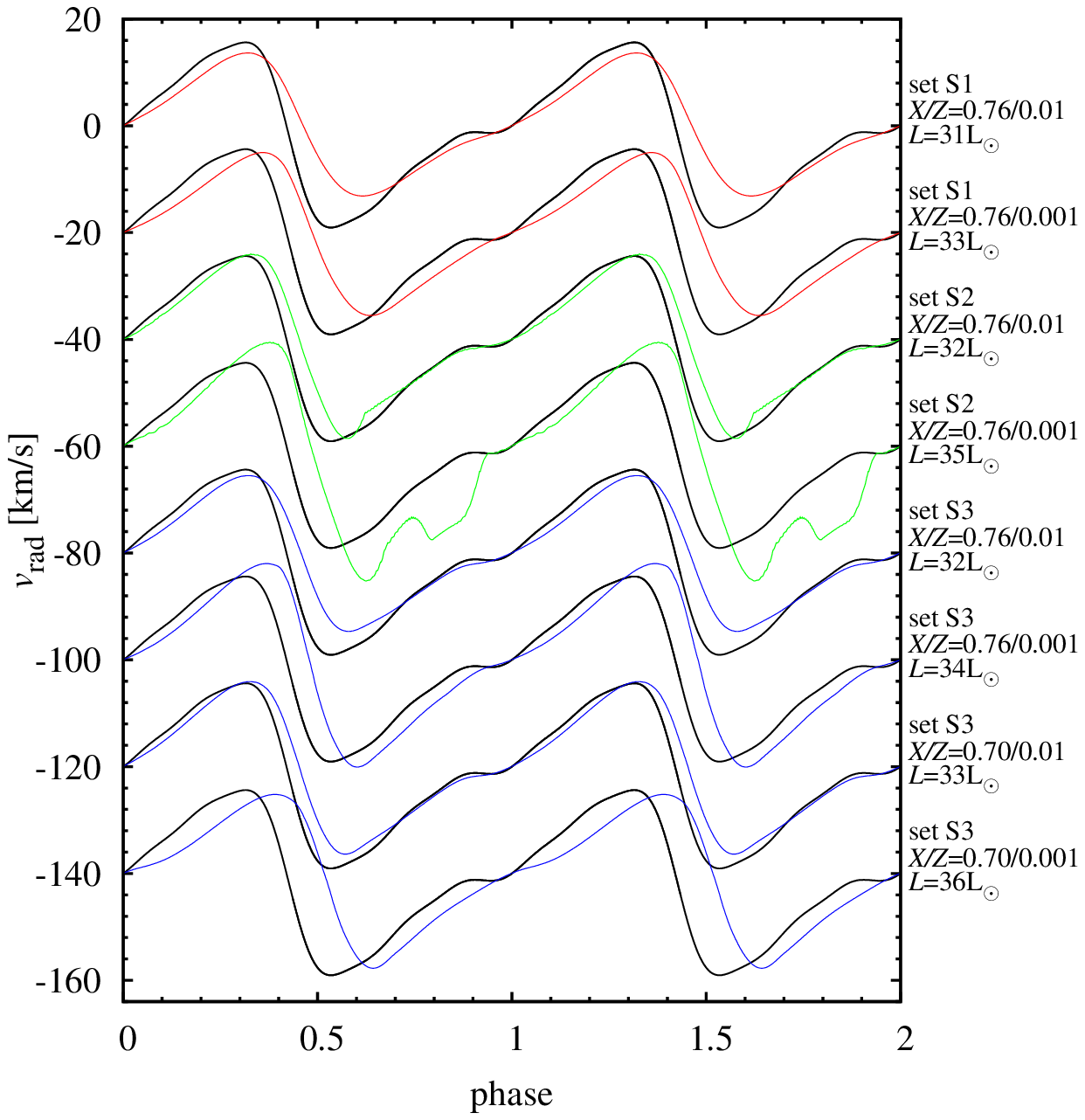}}
\caption{$I$-band (left panel) and radial velocity (right panel) curves for hydrodynamic models reproducing the observations of \BEP\ (black solid line, Fourier fit from Fig.~\ref{fig.lrvc}) best. Models adopting convective parameters of sets \setF, \setB\ and \setA\ are plotted with red, green and blue lines, respectively.}
\label{fig.bestmodels}
\end{figure*}

For the final selection of the best model we use the offset $d_{\rm all}$ constructed from all low order Fourier decomposition parameters and amplitudes for both the $I$-band light curve and radial velocity curve (ten parameters). The model that minimises $d_{\rm all}$ for a given model sequence is considered the best one. Properties of our best matching models for all considered model sequences are given in Tab.~\ref{tab.bestmodels}. In the last row we provide the value of $d_{\rm all}$. In the HR diagrams in Fig.~\ref{fig.hr} location of the best models is indicated with diamonds (in relevant panels). In Figs.~\ref{fig.best_lc} and \ref{fig.best_rv} the best models are marked with larger symbols. In Fig.~\ref{fig.bestmodels} the light and radial velocity curves for our best models for \BEP\ are compared with observations. The light curves were shifted vertically so the mean brightness is zero. All the curves were phased in the same way -- phase $0$ corresponds to the maximum radii (radial velocity goes through $0$). No other shifts were applied to the curves (except the vertical shift between consecutive models applied for clarity).

\begin{table*}
\caption{Properties of the best hydrodynamic models for \BEP.}
\label{tab.bestmodels}
\centering
\begin{tabular}{l|rrrrrrrr}
\hline
 set         &   \multicolumn{2}{c}{\setF}  &  \multicolumn{2}{c}{\setB}  &  \multicolumn{4}{c}{\setA}\\
 $X/Z$    & $0.76/0.01$ &   $0.76/0.001$ &   $0.76/0.01$ &   $0.76/0.001$ &  $0.76/0.01$ &   $0.76/0.001$ &  $0.70/0.01$ &   $0.70/0.001$ \\       
\hline
$L/\LS$     &   31   &  33    & 32     & 35     & 32     &  34    & 33 & 36\\
$\teff$ [K] & 6864.2 & 6913.8 & 6910.5 & 7003.0 & 6911.4 & 6957.3 & 6966.7 & 7053.6 \\
$R/\RS$     &  3.88  &  3.94  &  3.88  & 3.95   & 3.88   &  3.95  & 3.88 & 3.96\\
$P_2/P_0$   &  0.530 &  0.537  & 0.536 & 0.545   & 0.532  &  0.539 & 0.532 & 0.542 \\
$\Delta\varphi_1$ & -0.14 & -0.11 & -0.16 & -0.14 & -0.13 & -0.03  & -0.07& -0.06\\
\hline
 $d_{\rm all}$   & 0.597 &  0.421 & 0.382  & 0.742  & 0.290 & 0.630 & 0.200 & 0.227\\
\hline
\end{tabular}
\end{table*}

Analysis of Tab.~\ref{tab.bestmodels} and Figs.~\ref{fig.hr}, \ref{fig.best_lc}, \ref{fig.best_rv} and \ref{fig.bestmodels} allows to draw some interesting conclusions. 

({\it i}) All the best fitting models are located close to the blue edge of the instability strip (Fig.~\ref{fig.hr}). The effective temperatures and absolute luminosities lie in a narrow ranges of $\sim 6850-7050$\thinspace K and $31-36\LS$, respectively. More metal rich models ($Z=0.01$) are slightly ($50-100$\thinspace K) cooler and less luminous ($2-3\LS$) than metal poor models ($Z=0.001$). 

({\it ii}) Radius depends on metallicity and is equal to $\approx 3.88\RS$ for $Z=0.01$ and $\approx 3.95\RS$ for $Z=0.001$.

({\it iii}) Considering the $I$-band light curves (left panel of Fig.~\ref{fig.bestmodels}), clearly the best fit to the observed light curve is obtained for models with convective parameters of set \setA, and with $X=0.70$ (for both metallicities, bottom two curves in Fig.~\ref{fig.bestmodels}). These are the models with the lowest value of $d_{\rm all}$. A slightly worse fit is obtained for models with $X=0.76$. Among these models the best ones are those with convective parameters of sets \setB\ and \setA\ and with the higher metallicity ($Z=0.01$, third and fifth curve from top in Fig.~\ref{fig.bestmodels}). For all these  models the radial velocity curves well reproduce the observed curve, in particular, a bump on the ascending branch is present, although less pronounced than in observations. We conclude, that chemical composition typical for population I stars ($X=0.70$) seems more appropriate for \BEP\ (see discussion in Section~\ref{sec.concl}). 
 
({\it iv}) For the lower metallicity models ($Z=0.001$) a phase shift between the model and the observed light curves is apparent for models of all three sets of convective parameters (left panel in Fig.~\ref{fig.bestmodels}). A minimum brightness in these models follows the observed minimum by $\approx 0.1-0.15$ pulsation cycle. The shift is also present in the case of the radial velocity curves, for all the models. A phase of maximum expansion velocity in our models occurs later than is observed. Again the difference is slightly larger for the lower metallicity models. To explain these phase shifts we first note that we defined phase 0 as the phase of maximum radius. It is determined with the help of the radial velocity, which changes the sign. At around this phase a bump is present in the radial velocity curves, which manifests itself as a slower (less steep) variation of the radial velocity. Thus, its exact location may influence our phasing of the model curves. The location of the bump depends on the $P_2/P_0$ period ratio which is a function of metallicity, as is evident from an analysis of the values given in Tab.~\ref{tab.bestmodels}. At lower metallicity, the $P_2/P_0$ ratio becomes larger and the bump moves up the ascending branch of the radial velocity curve. The second effect is a systematic difference between the observed and the computed phase lag between the radial velocity curves and the light curves. Indeed, the phase lag of \BEP\ is positive and equal to $\Delta\varphi_1=0.055$, while the phase lags for all our models (Tab.~\ref{tab.bestmodels}) are systematically smaller and negative. A significantly better agreement is obtained for models with higher helium content, i.e. with $X=0.70$. We also note that in spite of the model phase lags being all negative, the phase of maximum expansion velocity always precedes the phase of maximum brightness, just as in the case of \BEP. The negative value of $\Delta\varphi_1$ results from the highly nonlinear shape of the light and radial velocity curves.

({\it vi}) The model for \BEP, that reproduces both the light and radial velocity variation best (has the lowest value of $d_{\rm all}$) is the model with convective parameters of set \setA\ and with chemical composition of population I stars ($X=0.70$, $Z=0.01$). Its absolute luminosity is $33\LS$ and effective temperature is $\approx 6970$\thinspace K. Among the models with chemical composition corresponding to population II stars, the best two models (with convective parameters of sets \setB\ and \setA) have the same metallicity ($Z=0.01$), luminosity ($L=32\LS$), and nearly the same effective temperature ($\approx 6910$\thinspace K). All our best matching models have very similar $P_2/P_0$ ratios, in agreement with interpretation that the bump is caused by the 2:1 resonance between the fundamental mode and the second overtone.

%%%%%%%%%%%%%%%%%%%%%%%%%%%%%%%%%%%%%%%%%%%%%%%%%%%%%%%%%%%%
\section{Non-Linear model survey for BEPs}\label{sec.survey}
%%%%%%%%%%%%%%%%%%%%%%%%%%%%%%%%%%%%%%%%%%%%%%%%%%%%%%%%%%%%

At first, \BEP\ was recognised as a classical RR~Lyrae star -- its light variation and radial velocity curves mimic that observed in RR~Lyrae stars pulsating in the fundamental mode. Only its presence in the eclipsing binary system allowed to determine its mass and to reveal its true nature. Its mass is only half the typical RR~Lyrae mass, too low to ignite helium. It cannot be a classical RR~Lyrae star, its a product of mass transfer during the binary evolution. The incidence rate of such stars estimated by \cite{gp12} is only $2$\thinspace per cent and only a fraction may be discovered through eclipses. Majority of such binaries are expected to be non-eclipsing systems. Consequently, direct estimate of mass of the pulsating component will not be possible. Then, the shape of the light/radial velocity curve may mimic that of RRab star, as it is the case for \BEP, and lead to wrong interpretation. The goal of this section is to conduct a large model survey of the light and radial velocity curves of BEPs, to check how to distinguish these stars from classical RRab stars and to check whether the case of \BEP\ -- RR~Lyrae impostor -- is a rule or an exception.

We consider model sequences with three different masses corresponding to one-third -- one-half the typical RR~Lyrae mass, $0.2\MS$, $0.26\MS$ (mass of \BEP) and $0.3\MS$, and with different absolute luminosities, $20\LS$, $25\LS$, $30\LS$, $35\LS$ and $40\LS$. Models are computed along horizontal stripes of constant luminosity with a step of $100$\thinspace K in effective temperature. Chemical composition is the same in all our models, $X=0.76$ and $Z=0.01$. Full model survey was computed adopting convective parameters of set \setF. In addition for 0.26\MS\ we have computed model sequences adopting convective parameters of sets \setB\ and \setA.

\begin{figure}
\centering
\resizebox{\hsize}{!}{\includegraphics{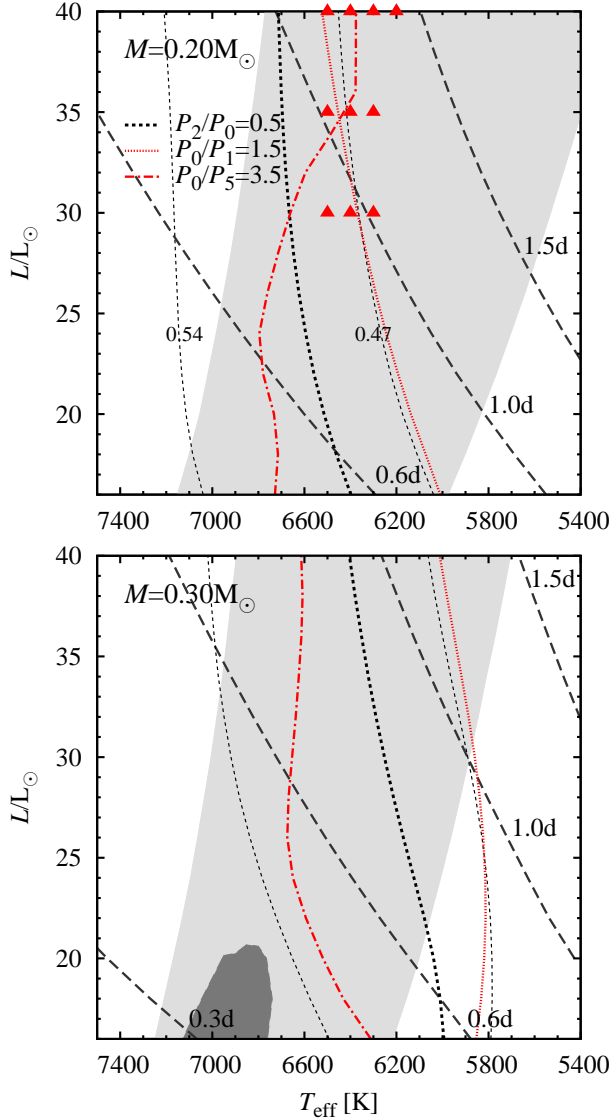}}
\caption{The HR diagrams for $0.20\MS$ (top panel) and $0.30\MS$ (bottom panel) models of set \setF\ and with $X=0.76$ and $Z=0.01$. Long-dashed lines are lines of constant fundamental mode period (0.3\thinspace d, 0.6\thinspace d, 1.0\thinspace d and 1.5\thinspace d). Loci of the 2:1 resonance with the second overtone and of the two half-integer resonances ($P_0/P_1=1.5$ and $P_0/P_5=3.5$) are also plotted. Triangles show the location of period-doubled models.}
\label{fig.hr2}
\end{figure}

In Fig.~\ref{fig.hr2} we present the HR diagrams for models with $M=0.20\MS$ (top panel) and $M=0.30\MS$ (bottom panel, adopting convective parameters of set \setF. For $M=0.26\MS$ the HR diagram is plotted in the top left panel in Fig.~\ref{fig.hr}. For the highest considered mass, the first overtone instability strip appears at the lowest luminosities. The models cover a wide period range from $0.3$ up to $1.5$ days. We focus our attention on models with $P_0$ below one day, as in RRab stars. As discussed in the previous section the 2:1 resonance is expected to shape the light and radial velocity curves of BEPs. For all considered masses, its loci (thick dotted lines in Figs.~\ref{fig.hr2} and \ref{fig.hr}) crosses the instability strip at fundamental mode periods between $0.6$ and $1$\thinspace days. 

\begin{figure}
\centering
\resizebox{\hsize}{!}{\includegraphics{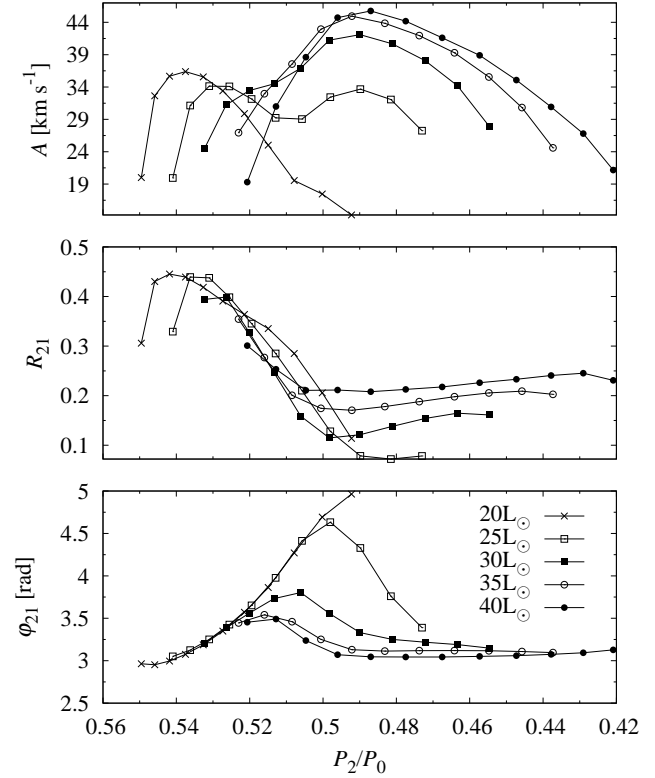}}
\caption{Peak-to-peak amplitude and the lowest order Fourier decomposition parameters for the radial velocity curves for $0.26\MS$ models (set \setF) of different luminosities (different symbols) plotted versus the linear $P_2/P_0$ period ratio. Pulsation period increases to the right.}
\label{fig.fou.HP}
\end{figure}

\begin{figure*}
\centering
\resizebox{.49\hsize}{!}{\includegraphics{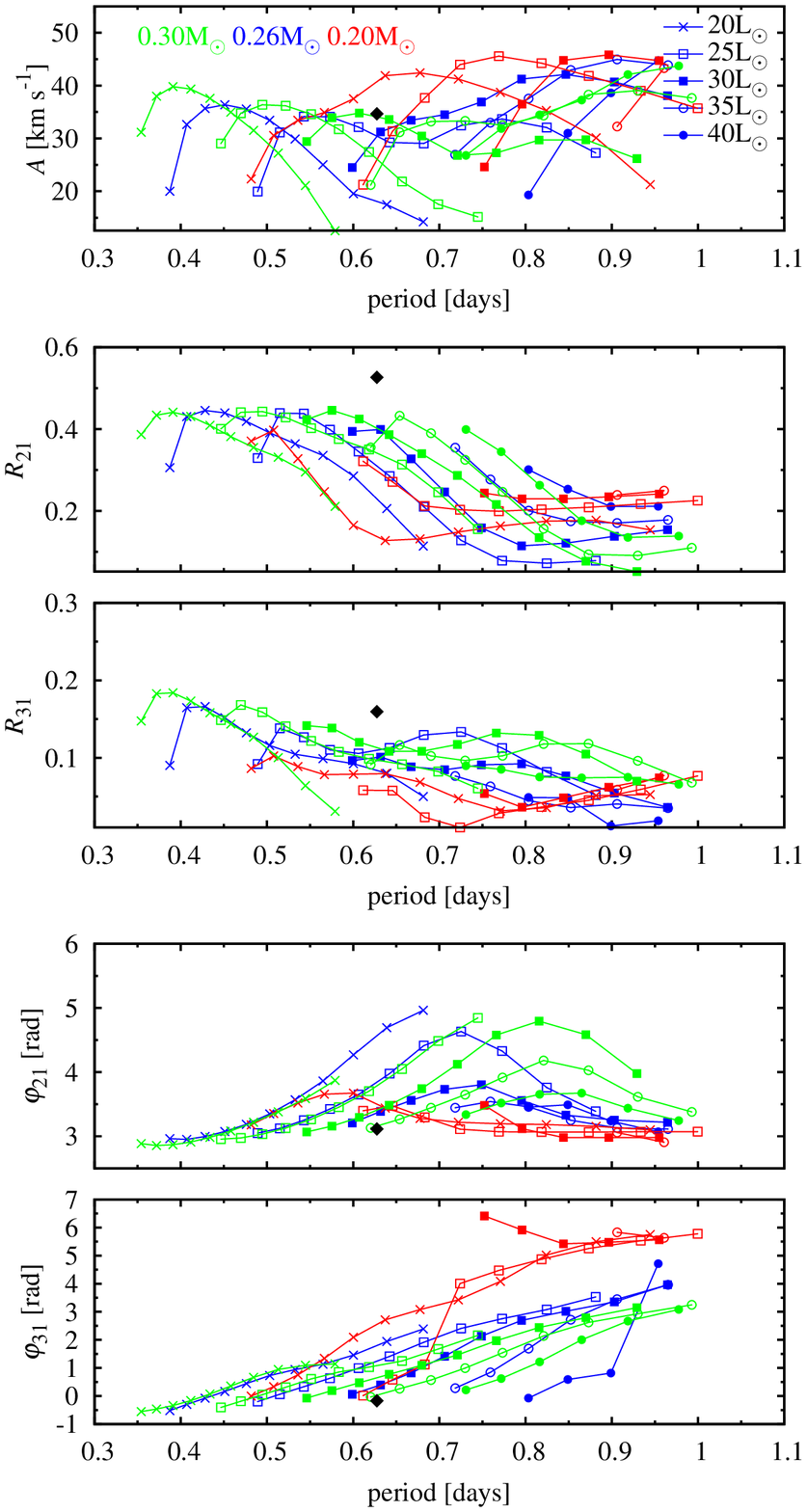}}
\resizebox{.49\hsize}{!}{\includegraphics{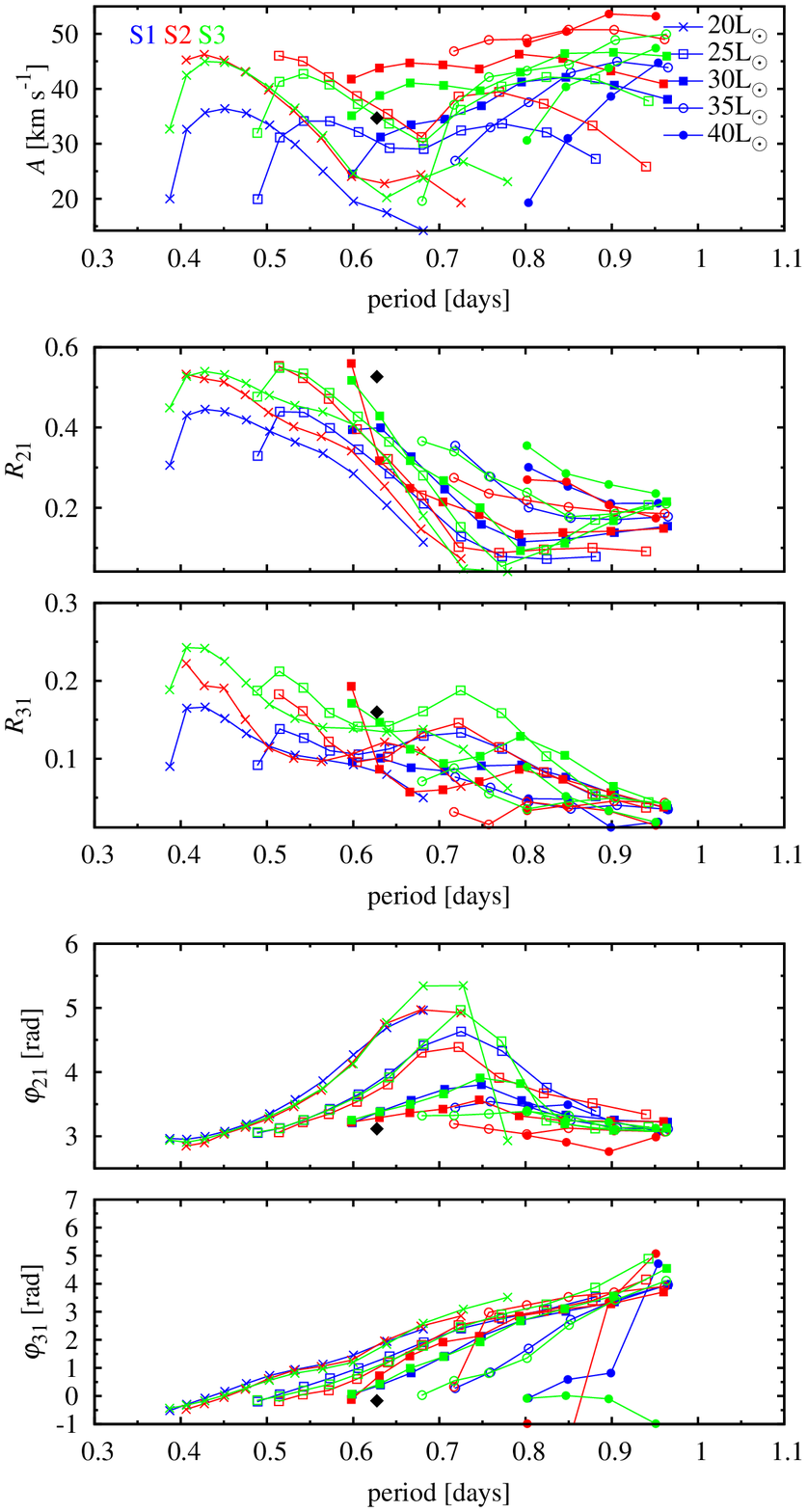}}
\caption{Fourier decomposition parameters for the radial velocity curves of non-linear models of BEPs. {\it Left panel:} models of different masses, $0.20\MS$, $0.26\MS$ and $0.30\MS$ along five stripes of different luminosity ($20-40\LS$). {\it Right panel:} Models of $0.26\MS$ and different luminosity and adopting different values of convective parameters (sets \setF, \setB\ and \setA). Diamond indicates the position of \BEP. Sine convention is used for Fourier phases.}
\label{fig.fou_freak_rv}
\end{figure*}

An analysis of the light and radial velocity curves and of the corresponding Fourier decomposition parameters clearly show that indeed, the 2:1 resonance with the second overtone causes the progression of the light and radial velocity curve's shape. This is illustrated based on model sequences with $M=0.26\MS$ (and different luminosities) and adopting convective parameters of set \setF. The bump progression, akin to that observed in classical Cepheids is best visible for radial velocity curves. The amplitude and the lowest order Fourier decomposition parameters are plotted in Fig.~\ref{fig.fou.HP} for model sequences with different luminosities. The Fourier parameters are plotted versus the $P_2/P_0$ period ratio. The progression of the Fourier coefficients clearly correlates with $P_2/P_0$ ratio and resembles very well the results obtained for classical Cepheid models \citep[e.g.][]{bmk90} and predictions based on analysis of amplitude equations for the 2:1 resonance \citep{kovb89}. A clear drop in the pulsation amplitude is observed, followed by the decrease and minimum of the $R_{21}$ ratio at somewhat larger values of $P_2/P_0$. In the case of the Fourier phase, a bell shape is apparent. The ascending part of the bell follows a universal progression -- $\varphi_{21}$ is a function of $P_2/P_0$ only (for $P_2/P_0>0.52$). The value of $P_2/P_0$ at which departure from this relation is observed and the highest possible value of $\varphi_{21}$ depends strongly on the luminosity. The correlation of the Fourier parameters progression with the $P_2/P_0$ ratio is lost if the Fourier parameters are plotted versus the period -- see the blue curves in the right panel of Fig.~\ref{fig.fou_freak_rv}, which represent the same model sequences as plotted in Fig.~\ref{fig.fou.HP}. No doubt, the 2:1 resonance is crucial in shaping the progression of the radial velocity curves. 

\begin{figure*}
\centering
\resizebox{.49\hsize}{!}{\includegraphics{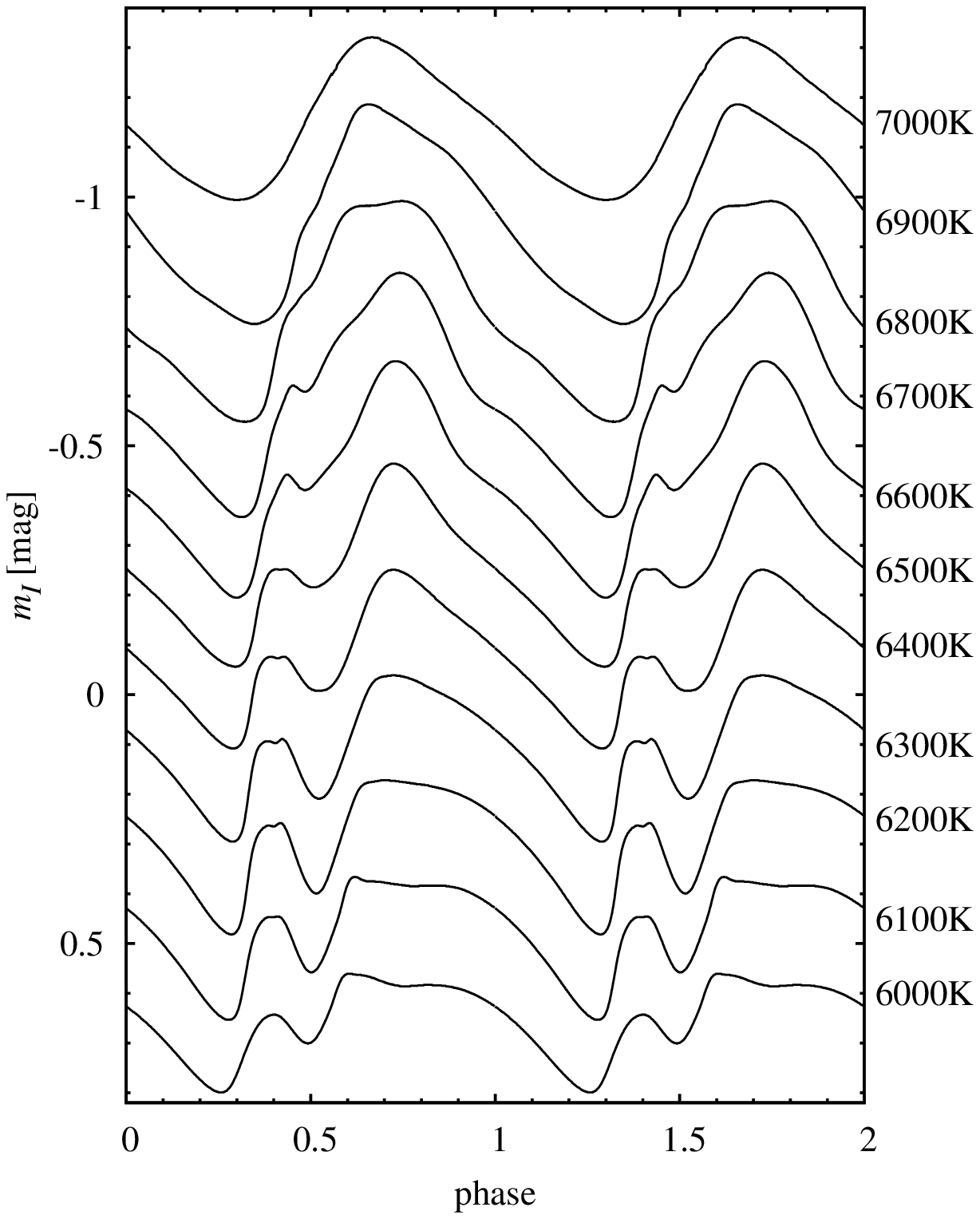}}
\resizebox{.49\hsize}{!}{\includegraphics{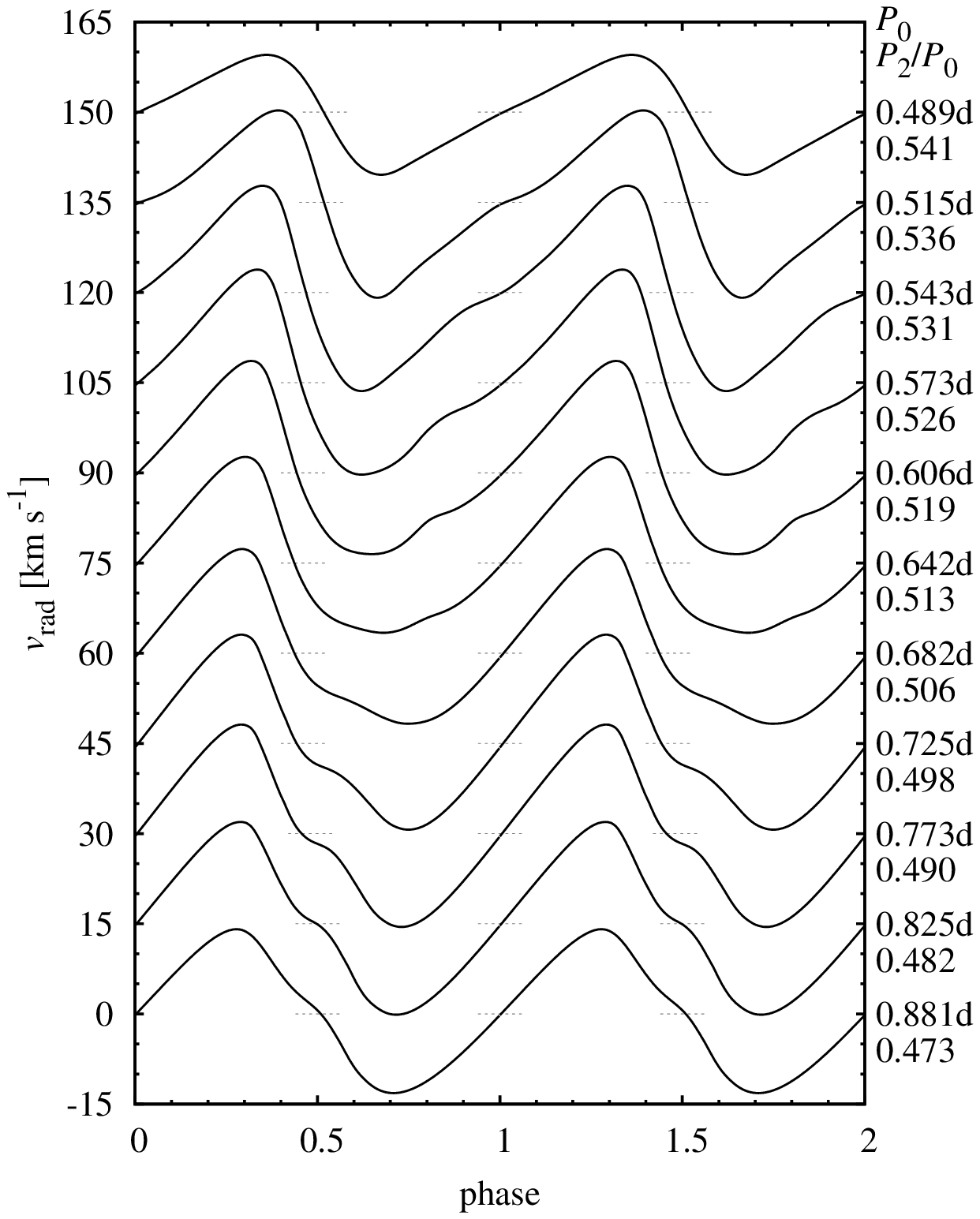}}
\caption{$I$-band light curves (left panel) and radial velocity curves (right panel) for hydrodynamic models of $0.26\MS$ computed along a line of constant luminosity, $L=25\LS$. Models adopt convective parameters of set \setF. Effective temperature is given on the right hand side of the $I$-band curves, fundamental mode period and linear $P_2/P_0$ period ratio is given on the right hand side of the velocity curves. For clarity, the consecutive light curves were shifted by $0.2$\thinspace mag and radial velocity curves by $15$\thinspace km\thinspace s$^{-1}$. In the latter case, short horizontal line segments mark the zero velocity level. Velocity curves were scaled by constant projection factor, $p=1.38$.}
\label{fig.HP_curves}
\end{figure*}

In Fig.~\ref{fig.HP_curves} (right panel) we plot the radial velocity curves along a sequence of $L=25\LS$. In contrast to Fig.~\ref{fig.PER.curves}, now a wide range of period ratios is probed, which clearly reveals the bump progression. The bump appears first in the middle of the ascending branch of the velocity curve, at the phase of maximum radius ($P_2/P_0\approx 0.54$). Then, as period increases and period ratio decreases, the bump moves towards the phase of maximum expansion velocity ($P_2/P_0\approx 0.51$). For period ratios below $\approx 0.5$ the bump is present on the descending branch and disappears for $P_2/P_0<0.47$. 

The bump progression is also visible for the $I$-band light curves (left panel of Fig.~\ref{fig.HP_curves}), although in this case the interpretation is more difficult, because of additional features present in the light variation. At the highest effective temperature the light curve is roughly triangular. A trace of bump appears on the descending branch, just after the maximum brightness, which next, moves towards the ascending branch and deforms the brightness maximum. For effective temperatures below $6700$\thinspace K ($P_2/P_0<0.52$) a pronounced bump is present on the ascending branch. It is present even for the lowest temperature model, and in general, is a feature of our models located in the centre and close to the red edge of the instability strip, even with period ratios as low as $P_2/P_0\approx 0.44$ (not shown in Fig.~\ref{fig.HP_curves}), although it is much less pronounced in such models. It is likely that other non-resonant effects play a role in shaping this bump. Another feature of the light curves, well visible in Fig.~\ref{fig.HP_curves} is a nearly flat and long phase of high brightness for cooler models  ($\teff\le 6200$\thinspace K). At the lowest effective temperatures, two brightness maxima at similar level are present. Origin of such shape is most likely non-resonant and connected to the development of extensive zone of efficient convective transfer at the phase of maximum compression, which is sustained then for nearly half the pulsation cycle.

Analysis of Figs.~\ref{fig.fou.HP} and \ref{fig.HP_curves} indicates that the 2:1 resonance influences the shape of the light and radial velocity curves in a wide range of period ratios, which we estimate to $0.47<P_2/P_0<0.54$. The border values of $P_2/P_0$ are plotted with thin dotted lines in Fig.~\ref{fig.hr2} and labelled with $P_2/P_0$ value. Clearly the 2:1 resonance is influential in a significant part of the instability strip and hence, is the most important effect shaping the light and radial velocity curves of BEPs.

\begin{figure}
\centering
\resizebox{\hsize}{!}{\includegraphics{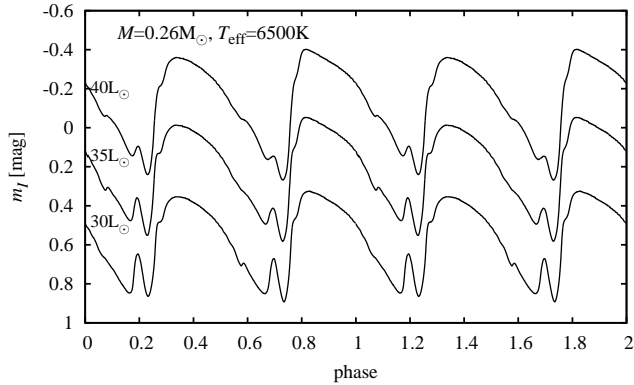}}
\caption{Exemplary $I$-band light curves for models with period-doubling effect.}
\label{fig.PD}
\end{figure}

Another resonant effect which we find in our models is a period-doubling effect. For some models, we observe alternating deep and shallow minima/maxima in the light and radial velocity variation -- see Fig.~\ref{fig.PD} for exemplary light curves. Consequently, pulsation repeats after two fundamental mode periods. The behaviour is found only for some model sequences and is restricted to rather narrow parameter ranges. For models adopting convective parameters of set \setF, period doubling was found only in the least massive, most luminous models. Their location is plotted with triangles in the top panel of Fig.~\ref{fig.hr2}.

As analysed by \cite{mb90} period doubling in the hydrodynamic models is caused by the half-integer resonances. The effect is observed in several types of large amplitude pulsators including RV~Tau variables \citep[caused by the 5:2 resonance, $P_0/P_2=2.5$,][]{mb90}, RR~Lyr stars showing the Blazhko effect \citep[caused by the 9:2 resonance, $P_0/P_9=4.5$,][]{szabo10} and in BL~Her stars \citep[caused by the 3:2 resonance, $P_0/P_1=1.5$,][]{sm12}. In Fig.~\ref{fig.hr2} we plot the loci of two half-integer resonances that might be influential in the case of BEPs models (red lines). These are the 3:2 resonance with the first overtone ($P_0/P_1=1.5$) and the 7:2 resonance with the fifth overtone ($P_0/P_5=3.5$). To cause the period doubling, the resonant overtone mode cannot be damped too strongly \citep{mb90}. This is the case for both resonances. The fifth overtone is a trapped envelope mode in the discussed models, for which the growth rate is high \citep{bk01}. Without the relaxation technique \citep[e.g.][]{stel74} however, it is hard to identify which resonance is operational in these models. As the models are limited to narrow parameter ranges we do not study this issue in more detail here.

We now turn to the discussion of the overall properties of the model light and radial velocity curves for BEPs, in terms of their Fourier decomposition parameters. To provide a feeling how reliable our non-linear code is in predicting the Fourier parameters, and to provide a background for discussing the parameters of BEPs, we have conducted a model survey for classical RR~Lyrae stars, briefly mentioned in Section~\ref{sec.hydro}, results of which were presented in Fig.~\ref{fig.fou_normF}. These models adopt the convective parameters of set \setF, but results for other sets of convective parameters are very similar. Amplitude and $R_{21}$ amplitude ratio are modelled satisfactorily, a somewhat too low values of $R_{31}$ are predicted. Fourier phases agree well with the observations, indicating that majority of RR~Lyrae stars have absolute luminosity around $50\LS$.

\begin{figure*}
\centering
\resizebox{.49\hsize}{!}{\includegraphics{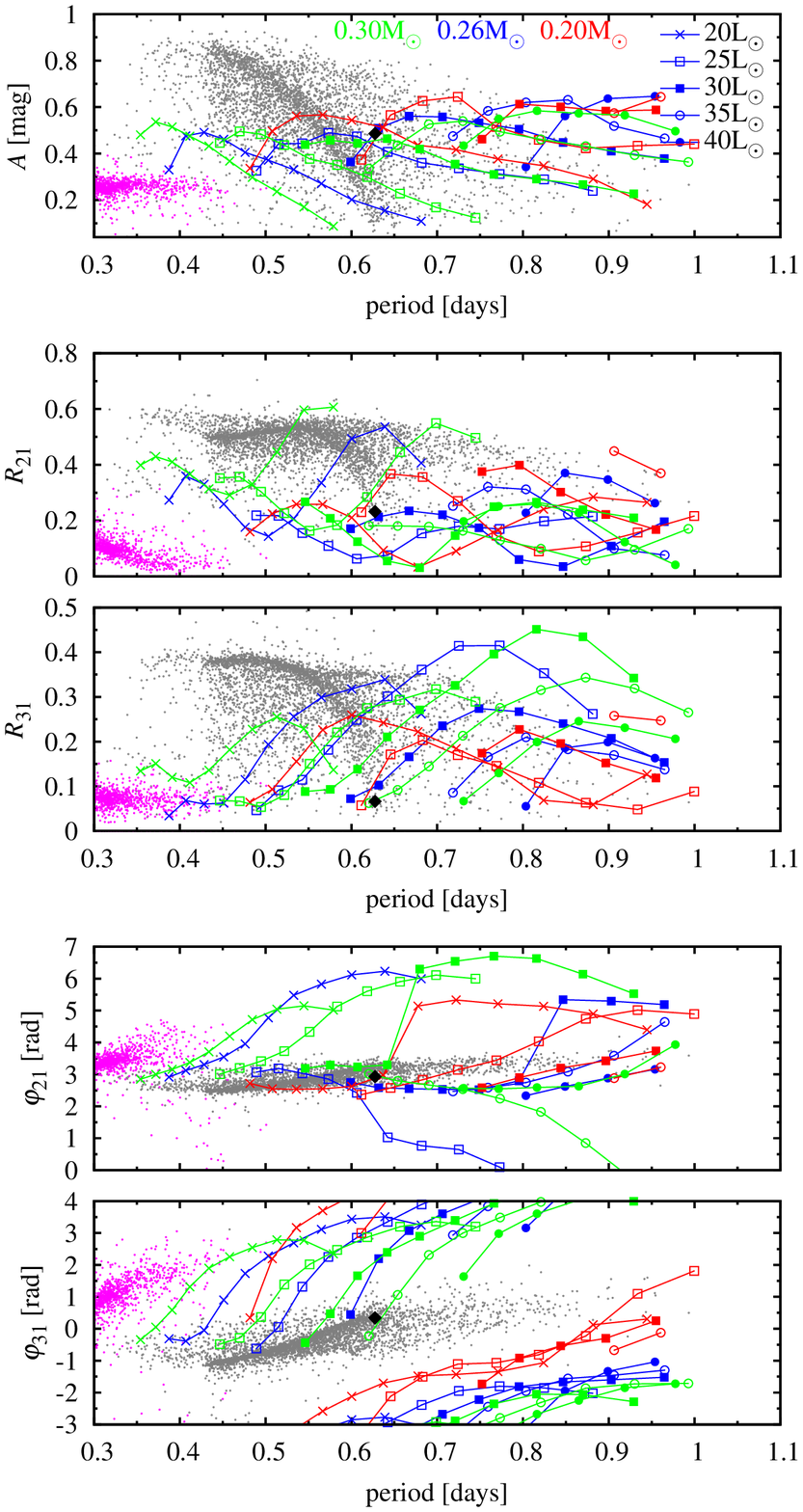}}
\resizebox{.49\hsize}{!}{\includegraphics{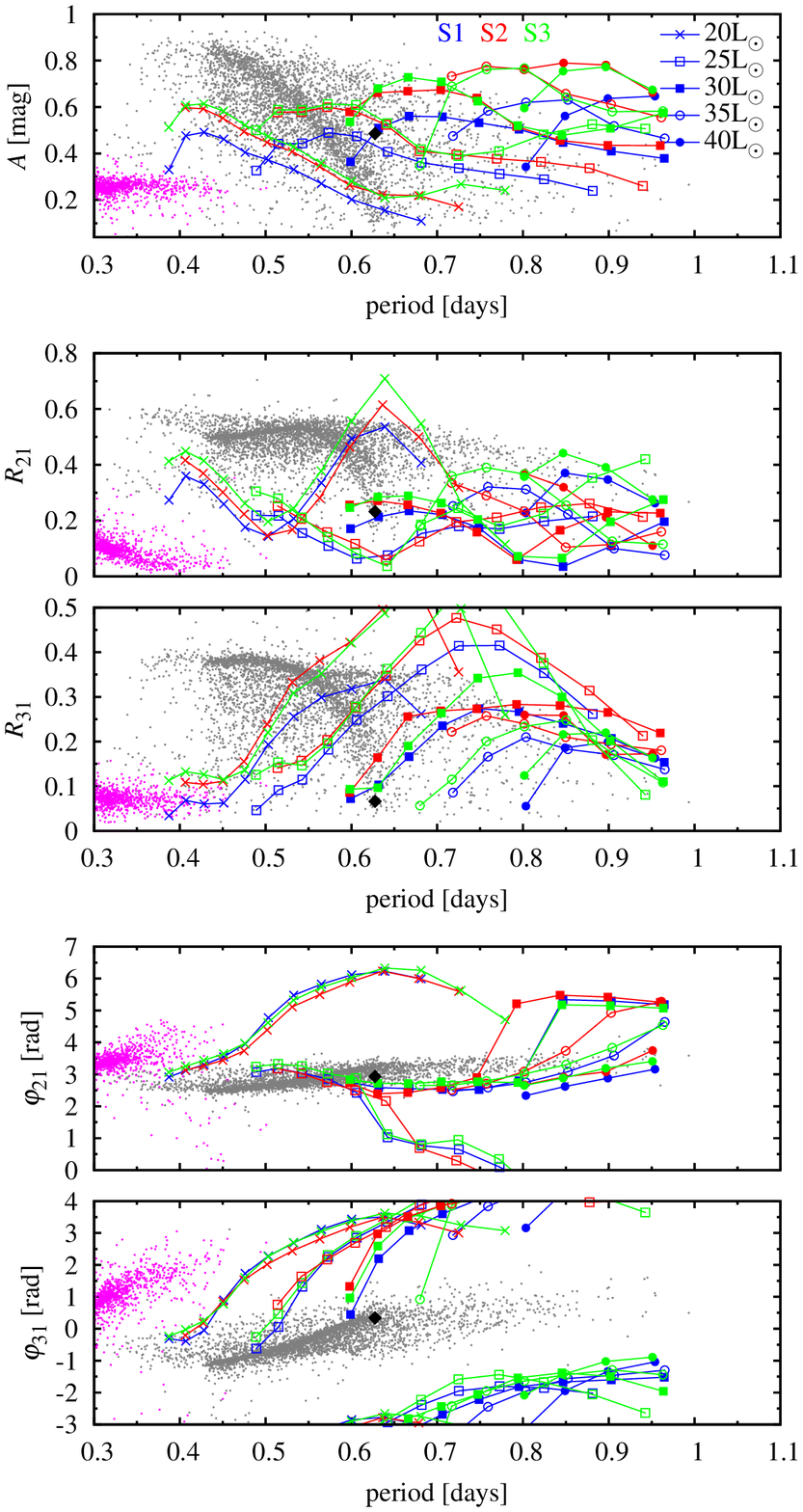}}
\caption{Fourier decomposition parameters for the $I$-band light curves of non-linear models of BEPs. {\it Left panel:} models of different masses, $0.20\MS$, $0.26\MS$ and $0.30\MS$ along five horizontal stripes with different luminosity ($20-40\LS$, different symbols). {\it Right panel:} Models of $0.26\MS$ and different luminosity and adopting different values of convective parameters (sets \setF, \setB\ and \setA). Diamond indicates the position of \BEP. For comparison, data for classical RR~Lyrae stars from the Galactic bulge from the OGLE catalogue are plotted \citep[grey dots for RRab stars, magenta dots for RRc stars, each third plotted;][]{sosz_rrl_blg}. Sine convention is used for Fourier phases.}
\label{fig.fou_freak_light}
\end{figure*}

Fourier parameters from similar survey of BEP models are plotted in Figs.~\ref{fig.fou_freak_light} ($I$-band light curves) and \ref{fig.fou_freak_rv} (radial velocity curves) for all considered model sequences. Left panels in these Figures illustrate the effects of different mass of the models (with convective parameters of the models fixed to that of set \setF), while right panels illustrate the effects of using different convective parameters in the modelling (sets \setF, \setB, and \setA), while mass of the models is fixed to $0.26\MS$. For each case, five model sequences with different absolute luminosities were computed. Successive models in each sequence were computed with a step of $100$\thinspace K in effective temperature. The first, shortest period model in each sequence is the hottest model located at the direct proximity of the blue edge of the instability strip. It is clear that with increasing luminosity, model sequences shift towards longer periods. In all the Figures we also mark the location of \BEP\ (diamond) and for the $I$-band light curves we also plot the location of RRab and RRc (first overtone pulsators) stars from the Galactic bulge for a reference. Such comparison is not possible in the case of radial velocity parameters. Only for a few RR~Lyrae stars coverage of radial velocity curves is good enough to allow determination of their Fourier parameters. Consequently, a meaningful comparison is not possible at the moment.

We first discuss the Fourier parameters for the $I$-band light curves displayed in Fig.~\ref{fig.fou_freak_light}. As discussed before, we expect that the 2:1 resonance will have a significant effect in shaping the light curves. We first note that progressions of Fourier parameters with pulsation period are very similar for different model sequences displayed in the Figure. For example the $R_{21}$ progressions have a wavy shape with local minima in the middle and at the edges of the progression. $R_{21}$ drops for the shortest and the longest period as a result of amplitude decrease at the edges of the instability strip. The decrease of $R_{21}$ in the middle of the progression is a resonant effect (see Fig.~\ref{fig.fou.HP}). As resonance centre falls at different periods for different model sequences, the minimum of $R_{21}$ also shifts in period. For example, for models with $M=0.20\MS$ the resonance centre is located at $P_0\approx 0.6$\thinspace days for $20\LS$ and at $P_0\approx 1$\thinspace day for $40\LS$ (top panel in Fig.~\ref{fig.hr2}).

The predicted peak-to-peak amplitudes of BEPs are below $0.6$\thinspace mag in the $I$-band, which is smaller than the typical amplitudes of RRab stars with periods shorter than $\approx 0.55$\thinspace days. In this period range also the predicted amplitude ratios are smaller than in RRab stars. We note however, that the predictions for amplitudes and amplitude ratios should be treated with caution. Amplitudes and consequently amplitude ratios strongly depend on the convective parameters adopted in the model, which is well visible if one compares the results for the models of sets \setF\ and \setA\ in the right panel of Fig.~\ref{fig.fou_freak_light}. The lower the eddy viscosity (set \setA), the higher become the amplitudes and amplitude ratios. Convective parameters should be calibrated to match the observational constraints, in particular the amplitude-period relation for a given group of pulsators. With only one BEP known at the moment the calibration is necessarily very rough. Fortunately, the Fourier phases do not depend strongly on the pulsation amplitude, which is clear from the analysis of the bottom right panels of Figs.~\ref{fig.fou_freak_light} and \ref{fig.fou_freak_rv}. The Fourier phases depend only weakly on the adopted convective parameters as well. Thus, the Fourier phases are the best and most reliable predictions of our modelling.

The progressions for the Fourier phases (bottom panels of Fig.~\ref{fig.fou_freak_light}) are very interesting. Note that for $\varphi_{31}$ the model sequences were plotted twice, shifted by $2\pi$. The $2\pi$ ambiguity in the values of the Fourier phases should be kept in mind while analysing the Figures. We note that $\varphi_{21}$ progressions strongly depend on the mass and on the absolute luminosity adopted in the models. Thus, $\varphi_{21}$ may provide useful constraints on the physical parameters of BEPs discovered in the future. Although for the majority of our models $\varphi_{21}$ significantly differs from the RRab progression, values typical for RRab stars are also possible, particularly for models located on the hot side of the instability strip. Consequently $\varphi_{21}$ is not the best indicator of BEP. It is the $\varphi_{31}$ Fourier phase that may be used to distinguish BEPs from RR~Lyrae stars. Clearly, the values of $\varphi_{31}$ predicted for BEPs differ significantly form the values typical for RR~Lyrae stars. Only very close to the blue edge of the instability strip (and this is the case for \BEP) the values of $\varphi_{31}$ are similar for the two groups. Through the majority of the instability strip $\varphi_{31}$ is larger in the BEPs models. For cooler models the $\varphi_{31}$ value increases. In the middle of the instability strip the difference between $\varphi_{31}$ as predicted for BEPs and the RRab progression is around $\pi$\thinspace rad. For lower masses the difference becomes smaller, but even for $0.2\MS$ models the difference is about $2$\thinspace rad. 

We also note that at shorter periods BEP models are also different from RRc stars. The predicted amplitudes are larger than those of RRc stars, and the same is true for the $R_{21}$ ratio. The modelled Fourier phases, both $\varphi_{21}$ and $\varphi_{31}$, fall in between the RRab and RRc progressions. The phases are larger than in RRab stars, but smaller than in RRc stars.

In the case of the radial velocity curves, the Fourier parameters displayed in Fig.~\ref{fig.fou_freak_rv} should be treated as predictions, to be compared with future observations of BEPs. It is hard to pinpoint the expected differences between the Fourier parameters for BEPs and those for RRab stars, as radial velocity observations of RRab stars are currently very scarce. The $\varphi_{21}$ values seem to be the most interesting ones for radial velocity curves. They depend only slightly on the adopted convective parameters. The bell shape is a result of the bump progression, and this is the most important difference between BEPs and RRab stars. No bump progression is expected in the latter case. The presence of a bump on a radial velocity curve is thus a good indicator for a BEP. The situation is more difficult in the case of the light curves, as additional features (bumps, shoulders) which may result for example from the shock propagating in the atmosphere \citep[e.g.][]{bs94}, are commonly present in RRab stars. We also note that the $\varphi_{21}$ values for the radial velocity curves might be used as an indicator of mass and absolute luminosity.

%%%%%%%%%%%%%%%%%%%%%%%%%%%%%%%%%%%%%%%%%%%%%%%%%%%%%%
\section{Discussions and conclusions}\label{sec.concl}
%%%%%%%%%%%%%%%%%%%%%%%%%%%%%%%%%%%%%%%%%%%%%%%%%%%%%%

In Section~\ref{sec.resultsBEP} we have conducted a model survey to reproduce the observed properties of \BEP. Our best model adopts the convective parameters of set \setA, and a chemical composition typical for population I stars, i.e. $X=0.70$ and $Z=0.01$. The model is located at the blue edge of the instability strip, its absolute luminosity is $33\LS$ and effective temperature is $\teff\approx 6970$\thinspace K. The model light curve fits the observed light variation very well. Good agreement was also obtained for the radial velocity curve (Fig.~\ref{fig.bestmodels}). We have reproduced the bump on the ascending branch of the radial velocity curve of \BEP, and traced its origin to the 2:1 resonance between the fundamental mode and the second overtone, $P_2/P_0=0.5$. 

The models with a chemical composition typical for population II stars ($X=0.76$) also reproduce the observations of \BEP\ well and have nearly the same luminosity and effective temperature. We note that although the mass of \BEP\ is very low and its pulsation resembles that of an RR Lyrae star, \BEP\ is the product of mass transfer in the binary system.  5.4\thinspace Gyr ago the progenitor of \BEP\ was a $1.4\MS$ star, as models of \cite{gp12} indicate and thus, the chemical composition typical for population I stars is very likely the correct one.

We also considered models with other values of convective parameters. The resulting properties of the best matching models for \BEP\ are then very similar to those quoted above (see Tab.~\ref{tab.bestmodels}) and therefore depend only slightly on the detailed treatment of convection in the model.

We note that the phase lag between the radial velocity curve and the light curve is systematically smaller in our models than that observed in \BEP. More computations are needed to find out how this discrepancy might be removed. The current models indicate that a slight helium enrichment might be needed. Also effects of turbulent convection, not considered in this study (e.g. turbulent pressure, overshooting) need to be checked. A more detailed modelling of \BEP\ is planned in the near future.

The effective temperature of \BEP\ estimated by \cite{gp12} is $\approx 350$\thinspace K larger ($\teff=7320\pm 160$\thinspace K) than in our best matching model. Consequently, also the absolute luminosity is higher ($L=46.2\pm 9.3\LS$). The estimates of effective temperature and absolute luminosity of \BEP\ not only differ from our results, but are also in conflict with stellar pulsation theory. They place \BEP\ outside the instability strip, as the analysis of Fig.~\ref{fig.hr} clearly indicates. This conclusion is independent of the convective parameters and chemical composition adopted in the model calculations. 

The estimate of the effective temperature of \BEP\ by \cite{gp12} was based on the assumed $T_2=5000$\thinspace K effective temperature of the secondary component. It is a typical temperature for a $1.67\MS$ giant, but it is not an actual measurement. Then, the effective temperature of the pulsating component of the binary was calculated using a temperature ratio ($T_1/T_2=1.464\pm 0.010$) obtained from the analysis of the light curve of the binary. To check how the assumption on $T_2$ affects the results we have redone the analysis of \cite{gp12} but adopting lower values of $T_2$. It turned out that a decrease of $T_2$ by $100$\thinspace K leads to a decrease of $T_1$ by $\approx 240$\thinspace K. Thus, only a slight decrease of the effective temperature of the secondary component, by $\approx 150$\thinspace K, brings the effective temperature of \BEP\ into agreement with the effective temperature resulting from our best models. The absolute luminosity is still somewhat too high ($L\approx 38\LS$) as compared to our models, which is caused by slightly different estimates of the radius.

The radius of \BEP\ estimated by \cite{gp12} is $R=4.24\pm 0.24\RS$, and, as we checked, is independent on the assumed value of $T_2$. The values inferred from our models (see Fig.~\ref{fig.hr} and Tab.~\ref{tab.bestmodels}) are lower by up to $1.5\,\sigma$ ($1.5\,\sigma$ for models with $Z=0.01$ and $1.2\,\sigma$ for models with $Z=0.001$). Multicolour photometry of the binary system may help to determine the radius of \BEP\ more accurately and this way tell us how severe the discrepancy is.
   
We have conducted a large non-linear model survey for the new class of variable stars. The aim was to provide the criteria to distinguish BEPs from classical RR~Lyrae pulsators. The resonant bump progression caused by the 2:1 resonance between the fundamental mode and the second overtone is the main factor expected to determine the shapes of the light and radial velocity curves of BEPs. This is not the case for classical RR~Lyrae stars. In the latter case, the 2:1 resonance falls close to the red edge of the instability strip and hence, no bump progression is observed in classical RR~Lyrae stars. Thus, the presence of a bump may indicate the binary evolution of the RR~Lyrae impostor. We note that the location of the bump on the radial velocity curve is a measure of the $P_2/P_0$ ratio, which provides an excellent constraint on the stellar models. In the case of the light curves, one may expect pronounced bumps on the ascending branch and long and flat, or long and double-peaked phases of high brightness (Fig.~\ref{fig.HP_curves}). In some of our models with the lowest $M/L$ ratio we find a period-doubling effect (Fig.~\ref{fig.PD}).

The other excellent discriminant of the new class of pulsating stars is the value of the $\varphi_{31}$ Fourier phase for the $I$-band light curve. Except the models located extremely close to the blue edge of the instability strip, the expected values of $\varphi_{31}$  differ significantly from the typical values for RRab stars. The values of $\varphi_{21}$ on the other hand, might be useful indicators of stellar mass and absolute luminosity. We also note that BEPs are expected to change (decrease) their periods very fast \citep{gp12}.

Our models clearly indicate that \BEP\, with its light and radial velocity curves strongly mimicking the typical RR~Lyrae curves, might be a rare case. For the majority of BEPs, significant departures from the observational properties of RR~Lyrae stars are expected.

\section*{Acknowledgments}
We dedicate this paper to the late Bohdan Paczynski (``Bep'' to his Warsaw and Lick colleagues) a  brilliant pioneer of stellar evolution within the confines of a Roche potential. Our colleague and mentor, who motivate many of us to take the difficult and challenging route of eclipsing binaries. We are grateful to Pawel Moskalik for helpful discussions and suggestions, reading and commenting the manuscript. Fruitful discussions with Wojciech Dziembowski are also acknowledged. We are grateful to the referee, Werner Weiss, for his helpful suggestions concerning the presentation of this paper. We gratefully acknowledge financial support for this work from the Chilean Center for Astrophysics FONDAP 15010003, and from the BASAL Centro de Astrofisica y Tecnologias Afines (CATA) PFB-06/2007. Support from the Ideas Plus program of Polish Ministry of Science and Higher Education, and the FOCUS and TEAM subsidies of the Foundation for Polish Science (FNP) is also acknowledged. KS acknowledges the financial support of the National Science Center through the grant DEC-2011/03/B/ST9/03299. Model computations presented in this paper were conducted on the psk computer cluster in the Copernicus Astronomical Centre, Warsaw, Poland.

%\bsp

\label{lastpage}

\end{document}